\documentclass[twocolumn,showpacs,amsmath,amssymb,bibnotes]{revtex4}
\pdfoutput=1
\usepackage{bm}
\usepackage{graphicx,epsfig,subfigure,afterpage,bm}
\usepackage{amsfonts}
\newcommand{\be}{\begin{equation}}
\newcommand{\ee}{\end{equation}}
\newcommand{\bea}{\begin{eqnarray}}
\newcommand{\eea}{\end{eqnarray}}
\newcommand{\gdot}{\dot{\gamma}}

\newcommand{\bw}{\begin{widetext}}
\newcommand{\ew}{\end{widetext}}

\newcommand{\xhat}{\vecv{\hat{x}}}
\newcommand{\yhat}{\vecv{\hat{y}}}
\newcommand{\zhat}{\vecv{\hat{z}}}

\newcommand{\vecv}[1]{\bm{{#1}}}
\newcommand{\tens}[1]{\bm{{#1}}}
\newcommand{\nablu}{{\bf \nabla}}

\newcommand{\vt}{\tilde{\vecv{v}}}
\newcommand{\st}{\tilde{\psi}}
\newcommand{\wt}{\tilde{\omega}}

\newcommand{\ldv}{L_{\rm DV}}
\newcommand{\lvi}{L_{\rm VI}}

\newcommand{\Dx}{\delta_x}
\newcommand{\Dy}{\delta_y}
\newcommand{\Dt}{\delta_t}

\newcommand{\tdv}{T_{\rm DV}}
\newcommand{\tvi}{T_{\rm VI}}

\begin{document}

\title{The need for inertia in nonequilibrium steady states of sheared binary fluids}
\author{Suzanne M. Fielding}
\email{suzanne.fielding@manchester.ac.uk}
\affiliation{School of Mathematics and Manchester Centre for Nonlinear Dynamics, University of Manchester, Oxford Road, Manchester M13 9PL, United Kingdom }

\date{\today}
\begin{abstract}
  We study numerically phase separation in a binary fluid subject to
  an applied shear flow in two dimensions, with full hydrodynamics.
  To do so, we introduce a mixed finite-differencing/spectral
  simulation technique, with a transformation to render trivial the
  implementation of Lees-Edwards sheared periodic boundary conditions.
  For systems with inertia, we reproduce the nonequilibrium steady
  states reported in a recent lattice Boltzmann study. The domain
  coarsening that would occur in zero shear is arrested by the applied
  shear flow, which restores a finite domain size set by the inverse
  shear rate. For inertialess systems, in contrast, we find no
  evidence of nonequilibrium steady states free of finite size
  effects: coarsening persists indefinitely until the typical domain
  size attains the system size, as in zero shear. We present an
  analytical argument that supports this observation, and that
  furthermore provides a possible explanation for a hitherto puzzling
  property of the nonequilibrium steady states with inertia.
\end{abstract}
\maketitle

%%%%%%%%%%%%%%%%%%%%%%%%%%%%%%%%%%%%%%%%%%%%%%%%%%%%%%%%%%%%%%%%%%%%%%%%%%%%%

\section{Introduction}
\label{sec:intro}

When an initially homogeneous mixture of two incompressible fluids (A
and B) undergoes a deep temperature quench into the spinodal regime,
it becomes unstable with respect to spatial fluctuations in the
composition field $\phi(\vecv{x},t)$. The mixture then phase separates
into well defined domains of A-rich and B-rich fluid, which, after a
rapid initial transient, attain local equilibrium within each domain.
The system remains globally out of equilibrium on long timescales,
however, due to the excess energy that resides in the interfaces
between the domains.  Here we consider the maximally symmetric case of
a 50:50 mixture of two mutually phobic fluids with matched viscosities
and densities.

Following the initial transient of their formation, the domains slowly
coarsen in time through the action of the surface tension in the
interfaces that separate them.  (For reviews of phase ordering
kinetics, see
Refs.~\cite{hohenberg77,bray-aip-43-357-1994,cates-fd--1-1999}.)  In
this way, the excess interfacial energy of the system progressively
relaxes towards its minimal equilibrium value. This coarsening process
proceeds through three distinct regimes that are successively
dominated by diffusive, viscous and inertial dynamics.  In the limit
of an infinite system size $\Lambda\to\infty$ (taken first), the
typical domain size perpetually increases without bound: the system
never globally equilibrates, even in the limit of infinite time
$t\to\infty$ (taken second).  For any finite system size, coarsening
is in practice eventually cutoff once the typical domain size $L(t)$
attains the system size $\Lambda$.

Besides these systems that remain out of equilibrium as they slowly
relax towards a fully phase separated state, another class of systems
comprises those that are continuously driven out of equilibrium by the
external application of a steady shear flow. In this paper, we
consider systems that are {\em both} undergoing phase separation {\em
  and} simultaneously subject to an applied shear flow.  They thus
combine both of the nonequilbrium features just described. The main
question that we address is whether shear interrupts domain coarsening
and restores a nonequilibrium steady state with a typical domain size
$L(\gdot^{-1})$ set by the inverse of the applied shear rate $\gdot$;
or whether coarsening persists indefinitely, up to the system size, as
in zero shear.

Despite previous
experimental~\cite{hashimoto-jcp-88-5874-1988,takahashi-jr-38-699-1994,chan-pra-43-1826-1991,krall-prl-69-1963-1992,hashimoto-prl-74-126-1995,beysens-pra-28-2491-1983,chan-prl-61-412-1988,matsuzaka-prl-80-5441-1998,hobbie-pre-54-R5909-1996,qiu-pre-58-R1230-1998,lauger-prl-75-3576-1995},
numerical~\cite{shou-pre-61-R2200-2000,zhang-jcp-113-8348-2000,yamamoto-pre-59-3223-1999,corberi-prl-81-3852-1998,corberi-prl-83-4057-1999,corberi-pre-61-6621-2000,corberi-pre-62-8064-2000,rothman-prl-65-3305-1990,rothman-el-14-337-1991,chan-el-11-13-1990,qiu-jcp-108-9529-1998,padilla-jcp-106-2342-1997,ohta-jcp-93-2664-1990,wagner-pre-59-4366-1999,lamura-pamia-294-295-2001,lamura-epjb-24-251-2001,berthier-pre-6305--2001}
and
theoretical~\cite{onuki-jpm-9-6119-1997,lamura-pamia-294-295-2001,corberi-prl-81-3852-1998,corberi-prl-83-4057-1999,corberi-pre-61-6621-2000,corberi-pre-62-8064-2000,bray-ptrslsapes-361-781-2003,cavagna-pre-62-4702-2000,bray-jpag-33-L305-2000,Onuk97,doi-jcp-95-1242-1991}
work, this question remained open until the recent simulation studies
of Stansell et al. in Refs~\cite{stansell-prl-96--2006,cates-3d-2007}.
Using Lattice Boltzmann techniques, they gave convincing evidence for
the formation of nonequilibrium steady states, with domains of a
finite size set by the inverse shear rate, independent of the system
size. The domain morphology inherits the anisotropy of the applied
shear flow, and is therefore characterised by two lengthscales
$L_{\|}$ and $L_\bot$, respectively describing the major and minor
domain axes. A remarkable achievement of
Refs.~\cite{stansell-prl-96--2006,cates-3d-2007} was the exploration,
by judicious parameter steering, of a range of inverse shear rates
spanning six decades on a scaling plot. For reasons discussed below
this range would, a priori, be expected to encompass two different
regimes, separately dominated by viscous and inertial dynamics.
Surprising, therefore, is the observation of an apparently single
scaling with shear rate for each of $L_{\|}$ and $L_\bot$, across all
six decades.

\begin{figure*}[tbp]
  \includegraphics[scale=0.35]{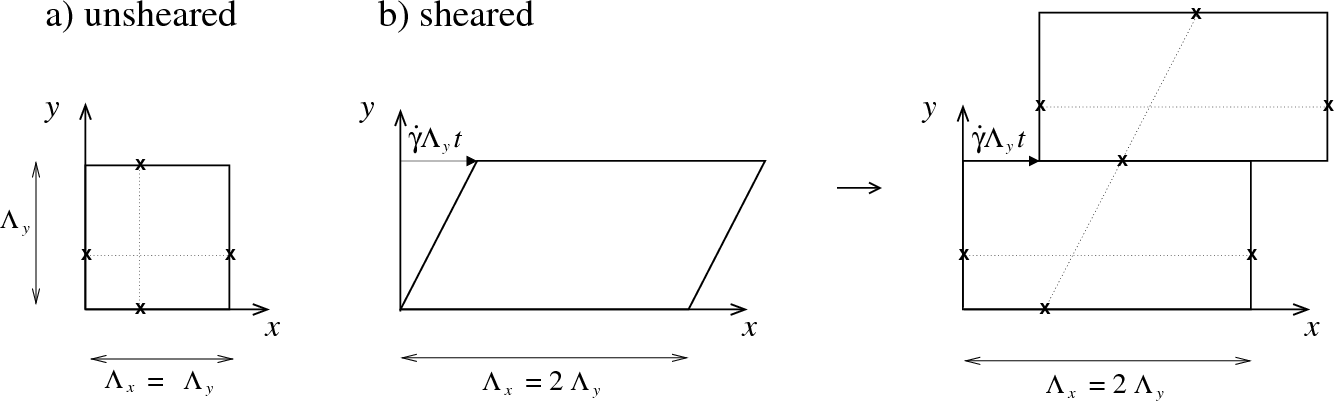}
  \caption{Flow geometries in the (a) unsheared and (b) sheared cases.
    The crossed connected by thin dotted lines show representative
    pairs of points connected by periodic boundary conditions.}
\label{fig:domain}
\end{figure*}

All the simulations reported in
Refs.~\cite{stansell-prl-96--2006,cates-3d-2007} have non zero fluid
inertia.  Indeed, even when attempting to access the limit of zero
inertia, a small but non-zero inertia remains a practical requirement
of the Lattice Boltzmann technique. In this paper, we investigate
phase separation under shear in systems that are strictly inertialess.
To do so, we apply a different simulation technique to this problem,
comprising finite differencing combined with Fourier spectral methods.
We also use a convenient transformation to render trivial the
implementation of sheared periodic boundary
conditions~\cite{onuki-jpsj-66-1836-1997}.

Our main numerical result will be that, in truly inertialess systems,
coarsening persists indefinitely, up to the system size, despite the
external shear flow. We will also present an analytical argument that
supports this observation. A simple extension to the same argument
provides a possible explanation for the existence of a single scaling
for each of $L_{\|}, L_\bot$ across all six decades of scaled shear
rate in the simulations of Ref.~\cite{stansell-prl-96--2006}, with
inertia.

We start by introducing the model equations, flow geometry and choice
of units in Secs.~\ref{sec:equations},~\ref{sec:geometry}
and~\ref{sec:units} respectively. In Sec.~\ref{sec:numerics} we
briefly outline our numerical method, of which further details are
given in the Appendices. The length and timescales that characterise
demixing are presented in Sec.~\ref{sec:lengthscales}, leading to a
discussion in Sec.~\ref{sec:parameters} of the choice of parameter
values in our simulations. In Sec.~\ref{sec:results} we present our
numerical results, starting in Sec.~\ref{sec:zeroShear} with the case
of coarsening in zero shear, before considering an applied shear flow
with and without inertia in Secs.~\ref{sec:inertia}
and~\ref{sec:noInertia} respectively.  Sec.~\ref{sec:link} contains a
linking discussion. The results in Sec.~\ref{sec:noInertia} are (we
believe) novel. In contrast, our aim in Secs.~\ref{sec:zeroShear}
and~\ref{sec:inertia} is simply to reproduce the behaviour seen in
previous lattice Boltzmann
studies~\cite{wagner-prl-80-1429-1998,kendon-jfm-440-147-2001,stansell-prl-96--2006},
thereby gaining confidence in our own simulation method. In
Sec.~\ref{sec:argument} we present an analytical argument that
supports our numerical observations, as well as those of
Ref.~\cite{stansell-prl-96--2006}.  Sec.~\ref{sec:conclusions} contains
a summary and an outlook to future work.

\section{Governing equations}
\label{sec:equations}

The fluid velocity field $\vecv{v}(\vecv{x},t)$ and pressure
$p(\vecv{x},t)$ are governed by the Navier-Stokes equation
\be
\label{eqn:ns}
\rho\left(\partial_t\, \vecv{v} + \vecv{v}\cdot \nablu \, \vecv{v} \right)=\eta \nabla^2 \vecv{v} -\phi \nablu \mu - \nablu p
\ee
together with the incompressibility constraint
\be
\label{eqn:incomp}
\nablu\cdot \vecv{v}=0.
\ee
Here $\rho$ is the fluid density and $\eta$ the viscosity.
$p(\vecv{x},t)$ is the thermodynamic part of the pressure tensor and
$\mu(\vecv{x},t)$ is a chemical potential, defined below.  In what
follows, we shall consider two dimensional (2D) flow in the $x-y$
plane with velocity components $v_x,v_y$.

To eliminate the pressure, we take the curl of Eqn.~\ref{eqn:ns}. The
$z$ component of the resulting equation is
\be
\label{eqn:vorticity}
\rho\left(\partial_t\,\omega + \vecv{v}\cdot\nablu \, \omega\right)=\eta
\,\nabla^2 \omega - \left[\nablu \wedge \phi \nablu \mu\right]\cdot \zhat
\ee
in which the vorticity $\omega$ obeys
\be
\nabla\wedge \vecv{v}=\omega \zhat.
\ee
To ensure that the incompressibility constraint~(\ref{eqn:incomp}) is
automatically obeyed, we define a streamfunction $\psi$ via
\be
\label{eqn:stream}
\vecv{v}=\nablu\wedge \psi \zhat,
\ee
which is related to the vorticity as follows:
\be
\label{eqn:streamVort}
\omega=-\nabla^2\psi.
\ee

The dynamics of the compositional order parameter $\phi(\vecv{x},t)$
are prescribed by an advection-diffusion equation of Cahn-Hilliard
type (see Refs.~\cite{bray-aip-43-357-1994,swift-pre-54-5041-1996})
\be
\label{eqn:phi1}
\partial_t \phi + \vecv{v}\cdot \nablu \phi = M\, \nabla^2 \mu.
\ee
Here $M$ is the mobility, assumed constant, controlling the rate of
intermolecular diffusion. The chemical potential
\be
\label{eqn:mu1}
\mu= G\,\phi(\phi^2-1)-\kappa \nabla^2\phi,
\ee
in which $G$ is a positive constant with the dimensions of stress.
$\kappa$ is also a positive constant, controlling the characteristic
width $l$ of the interface between domains:
\be
l=\left(\frac{\kappa}{G}\right)^\frac{1}{2}.
\ee
The surface tension of the interface is given by
\be
\sigma=\frac{2\sqrt{2}lG}{3}.
\ee

\section{Flow geometry}
\label{sec:geometry}

In the absence of shear we consider a square domain of size
$\Lambda_x=\Lambda_y$ in the $x-y$ plane (Fig.~\ref{fig:domain}a). We
adopt periodic boundary conditions in each direction:
\bea
\phi(x=0,y)&=&\phi(x=\Lambda_x,y)\;\;\;\forall\;\;\;y,\\
\phi(x,y=0)&=&\phi(x,y=\Lambda_y)\;\;\;\forall\;\;\;x,
%v_\alpha(x=0,y)&=&v_\alpha(x=\Lambda_x,y) \;\;\;\forall\;\;\;y,
\eea
and similarly for $v_x,v_y,\psi$ and $\omega$.

To study shear, we consider a system that is rectangular with
$\Lambda_x=2 \Lambda_y$. Conceptually, a shear flow of rate $\gdot$ is
applied by moving the upper boundary to the right at a constant speed
$\gdot \Lambda_y$, as shown in Fig.~\ref{fig:domain}b (left), so that
the velocity field then has a constant affine part $\gdot y\xhat$ and
an additive fluctuating contribution $\vt(\vecv{x},t)$. In practice,
this is implemented via Lees-Edwards sliding periodic boundary
conditions (Fig.~\ref{fig:domain}b, right), such that the system has
no physical borders but instead comprises an infinitely repeating
array of images arranged in horizontally sliding layers.  The boundary
conditions are then
\bea
\phi(x=0,y)&=&\phi(x=\Lambda_x,y)\;\;\;\forall\;\;\;y,\\
\phi(x,y=0)&=&\phi(x+\gdot
\Lambda_y t,y=\Lambda_y)\;\;\;\forall\;\;\;x.  \eea
These also apply to $\tilde{v}_x,\tilde{v}_y$, as well as the
corresponding fluctuating contribution to the streamfunction
$\tilde{\psi}$ and vorticity $\tilde{\omega}$.

\section{Choice of units}
\label{sec:units}

The model equations contain five parameters: $\rho, \eta, M, G$ and
$l$. (We must specify two of $G,l,\kappa$ and $\sigma$, and here have
chosen $G$ and $l$.) In addition, we must also specify the system's
dimensions $\Lambda_x,\Lambda_y$ and the applied shear rate $\gdot$,
making eight parameters in all. Three of these can be eliminated by
choosing appropriate units for length, time and stress. Accordingly,
we measure lengths in units of the vertical system size
$\Lambda_y\equiv 1$; stresses in units of $G\equiv 1$; and times in
units of the characteristic ``microscopic'' time for diffusion across
an interface,
\be
\tau_0=\frac{(\sqrt{2}l)^3}{3M\sigma}\equiv 1.
\ee
In these units, the governing equations comprise
Eqns.~\ref{eqn:vorticity},~\ref{eqn:stream} and~\ref{eqn:streamVort}
(unchanged) together with
\be
\label{eqn:phi}
\partial_t\, \phi + \vecv{v}\cdot \nablu \phi = l^2 \, \nabla^2 \mu
\ee
and
\be
\label{eqn:mu}
\mu= \phi(\phi^2-1)- l^2 \nabla^2\phi.
\ee
This leaves the rescaled density $\rho$, viscosity $\eta$, interfacial
width $l$, aspect ratio $\Lambda_x$ and applied shear rate $\gdot$ as
the five control parameters. In our numerical work we fix $\Lambda_x$
and $l$, varying $\rho,\eta$ and $\gdot$ between runs.

\section{Numerical method}
\label{sec:numerics}

In the laboratory frame, the numerical implementation of sheared
periodic boundary conditions is rather involved. We now discuss a
convenient transformation that renders it
trivial~\cite{onuki-jpsj-66-1836-1997}.  We do so here in outline
only, referring the interested reader to appendix~\ref{sec:appendix1}
for details.

The transformation comprises two steps. In the first, the velocity
field is expressed as the sum of a constant affine part and a
fluctuating contribution:
\be
\vecv{v}(\vecv{x},t)=\gdot y\xhat + \vt(\vecv{x},t).
\ee
In the second step, we transform to the cosheared frame
\be
(x,y,t)\to(x'=x-\gdot ty, y'=y,t'=t),
\ee
defining for convenience the cosheared gradient operator
\be
\label{eqn:coshearedGradient}
\nablu_c=\xhat\, \partial_{x'}+\yhat\,(-\gdot t\partial_{x'}+\partial_{y'}).
\ee
As shown in Appendix~\ref{sec:appendix1}, the final transformed
equation set, dropping the dashes for clarity, is then:
\be
\label{eqn:equation1}
{\omega}=-\nabla_c^2\,{\psi},
\ee
together with
\be
\label{eqn:equation2}
\rho\,\partial_{t}{\omega}+\rho\left(\partial_{y}{\psi}\partial_{x}{\omega}-\partial_{x}{\psi}\partial_{y}{\omega}\right)=\eta\,\nabla_c^2{\omega}-\left[\partial_{x}\phi\,\partial_{y} \mu-\partial_{y}\phi \,\partial_{x} \mu\right],
\ee
\be
\label{eqn:equation3}
\partial_{t}\phi+\left(\partial_{y}{\psi}\partial_{x}\phi-\partial_{x}{\psi}\partial_{y}\phi\right)=l^2\nabla_c^2\mu,
\ee
\be
\label{eqn:equation4}
\mu=\phi(\phi^2-1)-l^2\nabla_c^2\phi.
\ee
A priori, the bracketed expressions in Eqns.~\ref{eqn:equation2}
and~\ref{eqn:equation3} contain terms in the applied shear rate
$\gdot$. However in each bracket these are equal and opposite, and so
cancel. An equivalent statement is that the bracketed expressions are
invariant under shear.

This transformation considerably simplifies the implementation of an
applied shear flow. Indeed, looking at Eqns.~\ref{eqn:equation1}
to~\ref{eqn:equation4} we see that the only effect of shear is to
renormalise the gradient operator $\nabla\to\nabla_c$.  Furthermore,
the dynamical variables $\phi$, $\psi$ and $\omega$ are now subject to
{\em ordinary} periodic boundary conditions:
\bea
\phi(x=0,y)&=&\phi(x=\Lambda_x,y)\;\;\;\forall\;\;\;y,\nonumber\\
\phi(x,y=0)&=&\phi(x,y=\Lambda_y)\;\;\;\forall\;\;\;x,
\label{eqn:opbcs}
\eea
and similarly for $\psi$ and $\omega$.  This will allow us to use
Fourier transforms in our numerical algorithm, as described below.

Under the transformation described so far, the relative shear of the
laboratory and cosheared frames becomes large at long times, diverging
at a constant rate $\gdot$. This is clearly expected to give rise to
numerical instabilities. To circumvent this problem, once the relative
shear $s(t)$ attains $\Lambda_x/2 \Lambda_y$ we perform by hand an
instantaneous shift $s(t)\to s(t)-\Lambda_x/\Lambda_y$. In this way,
the function $s(t)$ is bounded between $s=-\Lambda_x/2 \Lambda_y$ and
$s=\Lambda_x/2 \Lambda_y$, comprising a sawtooth with sections of
constant slope $\gdot$ connected by negative step discontinuities of
height $\Lambda_x/\Lambda_y$ at equal time intervals $\Delta
t=\Lambda_x/\gdot\Lambda_y$. Because this shift involves moving the
upper wall by exactly one multiple of the system length, the periodic
boundary conditions (\ref{eqn:opbcs}) are unaffected. For times
between these shifts, the sole effect of this modification appears in
the cosheared gradient operator, which, in place of
(\ref{eqn:coshearedGradient}), is now
\be
\label{eqn:coshearedNabluFinal}
\nablu_c=\xhat\, \partial_{x}+\yhat\,(-s(t)\partial_{x}+\partial_{y}).
\ee
The special case of zero shear is trivially achieved by setting
$s(t)=0\;\forall\;t$.

In appendix~\ref{sec:appendix2}, we discuss the numerical algorithm
used to study the dynamical evolution of
$\phi(\vecv{x},t),\omega(\vecv{x},t)$ and $\psi(\vecv{x},t)$, as
specified by Eqns.~\ref{eqn:equation1} to~\ref{eqn:equation4}
and~\ref{eqn:coshearedNabluFinal}.  Readers who are not interested in
these issues can skip straight to Sec.~\ref{sec:lengthscales} without
loss of thread.

\section{Lengthscales and timescales}
\label{sec:lengthscales}

Following a deep temperature quench of an unsheared system into the
two phase regime, well defined domains of each phase form, separated
by sharp interfaces. See Fig.~\ref{fig:state_u}, for example, in which
the white (resp. black) patches are domains of A (resp.  B)-rich
fluid.  After a non-universal transient associated with their initial
formation, these domains progressively grow in size through the action
of surface tension, passing through regimes that are successively
dominated by diffusive, viscous and inertial dynamics.  During this
process of domain coarsening, the dynamical scaling
hypothesis~\cite{siggia,furukawa-aip-34-703-1985,bray-aip-43-357-1994} states that any structural
lengthscale $L(t)$ characterising the typical domain size ({\it e.g.},
as measured by some moment of the structure factor) should depend in
the same way on time $t$ as any other ({\it e.g.}, as measured by the
total amount of interface present in the system).  Within each regime,
dimensional analysis can be used to construct the functional form of
$L(t)$ out of the model parameters that are relevant to that
regime~\cite{siggia,furukawa-aip-34-703-1985}.
\begin{itemize}
\item In the diffusive regime, the system is unaware of the density
  $\rho$ and the viscosity $\eta$. Out of the remaining parameters
  $M,\sigma$, and the time $t$ we can construct only one lengthscale:
\be
\label{eqn:LD}
L_{\rm D}(t)=\left(M\sigma t\right)^{1/3}.
\ee
In our units $L_{\rm D}= \alpha\, l \,t^{1/3}$ with $\alpha=
(2\sqrt{2}/3)^{1/3}$.

\item In the viscous hydrodynamic regime, the system is unaware of $M$
  and $\rho$. From the remaining parameters $\eta,\sigma$ and $t$ we
  have
\be
\label{eqn:LV}
L_{\rm V}(t)=\frac{\sigma t}{\eta}.
\ee
In our units $L_{\rm V} = \alpha^3 l\, t/\eta$.

\item In the inertial hydrodynamic regime, the system is unaware of
  $M$ and $\eta$. From the remaining parameters $\rho,\sigma$ and
  $t$ we have
\be
\label{eqn:LI}
L_{\rm I}(t)=\left(\frac{\sigma t^2}{\rho}\right)^{1/3}.
\ee
In our units $L_{\rm I}= \alpha \left(l\, t^2/\rho\right)^{1/3}$.

\end{itemize}
By equating the growth laws $L_{\rm D} = L_{\rm V}$, we can construct
the characteristic lengthscale $\ldv$ at which we expect a crossover
from diffusive to viscous dynamics. Similarly by equating $L_{\rm
V}=L_{\rm I}$ we find the lengthscale $\lvi$ for crossover from
viscous to inertial dynamics.  Together with the ``microscopic''
interfacial width $l$ and the system size $\Lambda_x,\Lambda_y$, we
then have the following basic lengthscales
\begin{itemize}
\item The interfacial width $l$.

\item The characteristic lengthscale for crossover from diffusive to
  viscous hydrodynamic coarsening:
\be
\ldv=\sqrt{M \eta}.
\ee
In our units $\ldv=l \sqrt{\eta}$.

\item  The characteristic lengthscale for crossover from viscous
to inertial hydrodynamic coarsening:
\be
\lvi=\frac{\eta^2}{\rho \sigma}.
\ee
In our units $\lvi=3\eta^2/2\sqrt{2}\rho l$.

\item The system size $\Lambda_x,\Lambda_y$. In our units
  $\Lambda_y=1$ always.
\end{itemize}

Corresponding to these  are the following timescales
\begin{itemize}

\item The microscopic timescale for diffusion across
an interface
\be
\tau_0=\frac{2\sqrt{2}}{3}\frac{l^3}{M\sigma}=1.
\ee

\item  The characteristic timescale for crossover from
diffusive to viscous hydrodynamic coarsening 
\be
\tdv=\frac{\sqrt{M \eta^3}}{\sigma}=\frac{3}{2\sqrt{2}}\eta^{3/2}.
\ee

\item The characteristic timescale for crossover from viscous to
  inertial hydrodynamic coarsening 
\be
\tvi=\frac{\eta^3}{\rho\sigma^2}=\frac{9}{8}\frac{\eta^3}{\rho l^2}.
\ee

\item The characteristic time at which the domain size attains the
  system size in coarsening 
\be
T_{\rm system}.
\ee
\end{itemize}

%\vspace{0.4cm}
%In shear,  we have a further timescale set by
%
%\begin{itemize}
%\item the reciprocal applied shear rate $\gdot^{-1}$.
%\end{itemize}
%

So far we have discussed the growing structural lengthscale $L(t)$ in
general terms, without any specific definition.  In fact, there exist
many possible measures of the characteristic domain size. In zero
shear we use the following one
\be
\label{eqn:L_structure}
L_s=\left(\frac{\int dq_x\int dq_y S(\vecv{q})}{\int dq_x\int dq_y |q|^{-1}S(\vecv{q})}\right)^{-1},
\ee
defined via the structure factor $S(\vecv{q})$, which, as usual, is
the 2D Fourier Transform of $\phi(x,y)$.

The discussion so far in this section has related to unsheared
systems. Under an applied shear flow the domain morphology becomes
anisotropic. See Fig.~\ref{fig:steadyStates}, for example.
Accordingly, we consider the following
matrix~\cite{wagner-pre-59-4366-1999,stansell-prl-96--2006}
\be
\label{eqn:L_matrix}
D_{\alpha\beta}=\frac{l\int dx \int dy \,\partial_\alpha \phi \partial_\beta\phi}{\int dx \int dy \,\phi^2},
\ee
the reciprocal eigenvalues of which give us two lengthscales
$L_{\|},L_{\bot}$ characterising the long and short principal axes of
the domain morphology.

\begin{table*}
\centering
\subtable[Zero applied shear. Viscous hydrodynamic to inertial hydrodynamic regime.]{
\begin{tabular}{|p{1.3cm}|p{1.4cm}|p{1.5cm}|p{1.5cm}|p{1.0cm}|p{1.0cm}|p{1.0cm}|p{1.0cm}|p{1.2cm}|p{1.7cm}|p{2.0cm}|} \hline
Set  & $\eta$ & $\rho$ & $l$ & $\gdot$ & $\Lambda_x$ & $N_x$ & $N_y$ & $\Dt$ & $L_{\rm DV}$ & $L_{\rm VI}$  \\ \hline \hline
R028u & 0.111 & 136.0 & 0.00156 & 0.0 & 1.0 & 512 & 512 & 0.008 & 0.000519 & 0.0616  \\ 
R022u & 0.196 & 2510.0 & 0.00156 & 0.0 & 1.0 & 512 & 512 & 0.008 & 0.000690 & 0.0104  \\ 
R029u & 0.0467 & 893.0 & 0.00156 & 0.0 & 1.0 & 512 &512 & 0.008 & 0.000337 & 0.00166  \\
R020u & 0.0785 & 16180.0 & 0.00156 & 0.0 & 1.0 & 512 & 512 & 0.008 & 0.000437 & 0.000259 \\
R030u & 0.0122 & 6250.0 & 0.00156 & 0.0 & 1.0 & 512 & 512 & 0.008 & 0.000172 & 0.0000162 \\
R019u & 0.00876 & 32814.0 & 0.00156 & 0.0 & 1.0 & 512 & 512 & 0.008 & 0.000146 & 0.00000159 \\
R032u & 0.00391 & 20067.0 & 0.00156 & 0.0 & 1.0 & 512 & 512 & 0.008 & 0.0000975 & 0.000000518 \\
\hline
\end{tabular}
\label{tab:firsttable}
}
\qquad\qquad
\subtable[Zero applied shear. Diffusive to viscous hydrodynamic regime.]{        
\begin{tabular}{|p{1.3cm}|p{1.4cm}|p{1.5cm}|p{1.5cm}|p{1.0cm}|p{1.0cm}|p{1.0cm}|p{1.0cm}|p{1.2cm}|p{1.7cm}|p{2.0cm}|} \hline
Set  & $\eta$ & $\rho$ & $l$ & $\gdot$ & $\Lambda_x$ & $N_x$ & $N_y$ & $\Dt$ & $L_{\rm DV}$ & $L_{\rm VI}$  \\ \hline \hline
%DV1u & 1000.0 & 0.0 & 0.0025 & 0.0 & 1.0 & 512 & 512 & 0.016 & 0.0791 & $\infty$ \\
%DV2u & 100.0 & 0.0 & 0.0025 & 0.0 & 1.0 & 512 & 512 & 0.016 & 0.0250 & $\infty$ \\
%DV3u & 10.0 & 0.0 & 0.0025 & 0.0 & 1.0 & 512 & 512 & 0.016 & 0.00791 & $\infty$ \\
%DV4u & 1.0 & 0.0 & 0.0025 & 0.0 & 1.0 & 512 & 512 & 0.016 & 0.00250 & $\infty$ \\
%DV5u & 0.3 & 0.0 & 0.0025 & 0.0 & 1.0 & 512 & 512 & 0.016 & 0.00137 & $\infty$ \\
%DV6u & 0.1 & 0.0 & 0.0025 & 0.0 & 1.0 & 512 &512 & 0.016 & 0.000791 &  $\infty$ \\
DV1u & 1000.0 & 0.0 & 0.00156 & 0.0 & 1.0 & 512 & 512 & 0.016 & 0.0494 & $\infty$ \\
DV2u & 100.0 & 0.0 & 0.00156 & 0.0 & 1.0 & 512 & 512 & 0.016 & 0.0156 & $\infty$ \\
DV3u & 10.0 & 0.0 & 0.00156 & 0.0 & 1.0 & 512 & 512 & 0.016 & 0.00494 & $\infty$ \\
DV4u & 1.0 & 0.0 & 0.00156 & 0.0 & 1.0 & 512 & 512 & 0.016 & 0.00156 & $\infty$ \\
DV5u & 0.3 & 0.0 & 0.00156 & 0.0 & 1.0 & 512 & 512 & 0.016 & 0.000855 & $\infty$ \\
DV6u & 0.1 & 0.0 & 0.00156 & 0.0 & 1.0 & 512 &512 & 0.016 & 0.000494 &  $\infty$ \\
\hline
\end{tabular}
\label{tab:secondtable}
}
\qquad\qquad
\subtable[Applied shear, with inertia.]{        
\begin{tabular}{|p{1.3cm}|p{1.4cm}|p{1.5cm}|p{1.5cm}|p{1.0cm}|p{1.0cm}|p{1.0cm}|p{1.0cm}|p{1.2cm}|p{1.7cm}|p{2.0cm}|} \hline
Set  & $\eta$ & $\rho$ & $l$ & $\gdot$ & $\Lambda_x$ & $N_x$ & $N_y$ & $\Dt$ & $L_{\rm DV}$ & $L_{\rm VI}$  \\ \hline \hline
R028s & 0.111 & 136.0 & 0.00156 & 0.0765 & 2.0 & 1024 & 512 & 0.008 & 0.000519 & 0.0616  \\ 
R028bs & 0.444 & 8704.0 & 0.00156 & 0.0765 & 2.0 & 1024 & 512 & 0.008 & 0.00104 & 0.0154  \\ 
R022s & 0.196 & 2510.0 & 0.00156 & 0.0205 & 2.0 & 1024 & 512 & 0.008 & 0.000690 & 0.0104  \\ 
R022bs & 0.339 & 22590.0 & 0.00156 & 0.0355 & 2.0 & 1024 & 512 & 0.008 & 0.000908 & 0.00349  \\ 
R029s & 0.0467 & 893.0 & 0.00156 & 0.0341 & 2.0 & 1024 &512 & 0.008 & 0.000337 & 0.00166  \\
R029bs & 0.0809 & 8037.0 & 0.00156 & 0.0591 & 2.0 & 1024 &512 & 0.008 & 0.000444 & 0.000554  \\
R020s & 0.0785 & 16180.0 & 0.00156 & 0.0256 & 2.0 & 1024 & 512 & 0.008 & 0.000437 & 0.000259 \\
R030s & 0.0122 & 6250.0 & 0.00156 & 0.0410 & 2.0 & 1024 & 512 & 0.004 & 0.000172 & 0.0000162 \\
R019s & 0.00876 & 32814.0 & 0.00156 & 0.0251 & 2.0 & 1024 & 512 & 0.004 & 0.000146 & 0.00000159 \\
R032s & 0.00391 & 20067.0 & 0.00156 & 0.051 & 2.0 & 1024 & 512 & 0.004 & 0.0000975 & 0.000000518 \\
\hline
\end{tabular}
\label{tab:thirdtable}
}
\qquad\qquad
\subtable[Applied shear, without inertia.]{        
\begin{tabular}{|p{1.3cm}|p{1.4cm}|p{1.5cm}|p{1.5cm}|p{1.0cm}|p{1.0cm}|p{1.0cm}|p{1.0cm}|p{1.2cm}|p{1.7cm}|p{2.0cm}|} \hline
Set  & $\eta$ & $\rho$ & $l$ & $\gdot$ & $\Lambda_x$ & $N_x$ & $N_y$ & $\Dt$ & $L_{\rm DV}$ & $L_{\rm VI}$  \\ \hline \hline
%DV1s & 1000.0 & 0.0 & 0.0025 & 0.01 & 2.0 & 1024 & 512 & 0.016 & 0.0791 & $\infty$ \\
%DV2s & 100.0 & 0.0 & 0.0025 & 0.01 & 2.0 & 1024 & 512 & 0.016 & 0.0250 & $\infty$ \\
%DV3s & 10.0 & 0.0 & 0.0025 & 0.01 & 2.0 & 1024 & 512 & 0.016 & 0.00791 & $\infty$ \\
%DV4s & 1.0 & 0.0 & 0.0025 &  0.01 & 2.0 & 1024 & 512 & 0.016 & 0.00250 & $\infty$ \\
%DV5s & 0.3 & 0.0 & 0.0025 & 0.01  & 2.0 & 1024 & 512 & 0.016 & 0.00137 & $\infty$ \\
%DV6s & 0.1 & 0.0 & 0.0025 &  0.03 & 2.0 & 1024 &512 & 0.016 & 0.000791 &  $\infty$ \\
DV1s & 1000.0 & 0.0 & 0.00156 & 0.01 & 2.0 & 1024 & 512 & 0.016 & 0.0494 & $\infty$ \\
DV2s & 100.0 & 0.0 & 0.00156 & 0.01 & 2.0 & 1024 & 512 & 0.016 & 0.0156 & $\infty$ \\
DV3s & 10.0 & 0.0 & 0.00156 & 0.01 & 2.0 & 1024 & 512 & 0.016 & 0.00494 & $\infty$ \\
DV4s & 1.0 & 0.0 & 0.00156 &  0.01 & 2.0 & 1024 & 512 & 0.016 & 0.00156 & $\infty$ \\
DV5s & 0.3 & 0.0 & 0.00156 & 0.01  & 2.0 & 1024 & 512 & 0.016 & 0.000855 & $\infty$ \\
DV6s & 0.1 & 0.0 & 0.00156 &  0.03 & 2.0 & 1024 &512 & 0.016 & 0.000494 &  $\infty$ \\
\hline
\end{tabular}
\label{tab:fourthtable}
}
\caption{Parameter values used in simulation runs.}
\label{tab:master}
\end{table*}

\section{Parameter steering}
\label{sec:parameters}

As discussed in Sec.~\ref{sec:units}, the physical control parameters
that must be prescribed in any simulation run are the fluid density
$\rho$ and viscosity $\eta$, the interfacial width $l$, the applied
shear rate $\gdot$, and the system's aspect ratio $\Lambda_x$.
(Recall that we have set $\Lambda_y = 1$, $G = 1$ and $\tau_0 = 1$.)
We must also specify the number of numerical mesh points $N_x$ and
$N_y$, and the numerical timestep $\delta_t$.  We now discuss
appropriate choices of these parameter values that will allow us to
access the physical regimes of interest.

In the absence of an applied shear flow, $\gdot=0.0$, our aim is to
study coarsening of the isotropic domain morphology. See
Fig.~\ref{fig:state_u}, for example. In each of these runs, we
consider a square simulation box with $\Lambda_x=\Lambda_y =
\Lambda=1$, in our units.  In any given run, at any time $t$, our aim
is to ensure a separation of the four lengthscales
\be
\delta\ll l \ll L(t)\ll \Lambda.
\ee
Here $\delta=1/N$ is the mesh size, prescribed by the reciprocal of
the number $N=N_x=N_y$ of numerical mesh points in each spatial
dimension. Recall that $l$ is the width of the interface between
domains; $L(t)$ is the growing domain size; and $\Lambda\equiv 1$ the
system size.  We thereby restrict ourselves to physical lengthscales
$l$ and $L(t)$ that lie between the mesh and system size. To make this
window as large as possible, we take $\delta$ as small as possible,
using as many grid points $N^2=1/\delta^2$ as is numerically feasible.
The results presented below have $N=512$, checked for convergence to
the limit $N\to\infty$ at fixed $l$ against $N=1024$ (not shown). We
set the interfacial width $l$ to the minimum for which interfaces are
still fully resolved by this grid, taking $l=0.00156$, checked for
convergence to the limit $l/\Lambda
\to 0$ against $l=0.0025$ and $l=0.005$ (not shown).

In each simulation run, this leaves a window of approximate size $0.05
< L(t) < 0.25$ in which the domain size $L(t)$ is much larger than the
width $l$ of the interface between domains, and much smaller than the
system size $\Lambda=1$.  Accordingly, we are only able to study a
small piece of the full curve $L(t)$ in each run. In different runs,
therefore, we vary the viscosity $\eta$ and density $\rho$ to ensure
that we cover all regimes of interest.  As discussed in
Refs.~\cite{kendon-prl-83-576-1999,pagonabarraga-jsp-107-39-2002}, we
are then able to construct composite scaling curves $L(t)/\ldv$ and
$L(t)/\lvi$, each spanning several decades of scaled length and time.

Below we present results for two different series of runs.  In the
first, R028u to R032u in table~\ref{tab:firsttable}, we explore the
viscous and inertial hydrodynamic regimes, and the crossover between
them, by varying the crossover lengthscale $\lvi$. These are in fact
the parameter values used previously by Kendon and coworkers in their
lattice Boltzmann simulations~\cite{kendon-jfm-440-147-2001}, converted
into our units.
%Note that $\ldv\ll l$ in each case,
%ensuring negligible diffusion on the lengthscale of a macroscopic
%domain.

In the second series of runs, DV1u to DV6u in
table~\ref{tab:secondtable}, we explore the diffusive and (again)
viscous hydrodynamic regimes, and the crossover between them.
Accordingly, we set $\rho=0$ such that $\lvi$ is infinite and the
inertial regime is never attained.  Across the different runs we vary
the predicted crossover lengthscale $\ldv=l\sqrt{\eta}$ by sweeping
the viscosity in the range $\eta=10^{-2}\cdots \eta=10^4$. For the
largest viscosity values, the system explores the diffusive regime.
For the smallest values, the system passes into the viscous regime as
soon as well defined domains have formed, without any identifiable
regime of diffusive domain growth.

%All the runs in table~\ref{tab:master} have an interfacial width set
%by $l=0.00156$, following Ref. However we found these to be only
%marginally converged with respect to mesh fineness, for the maximum
%number of mesh points feasible within present computational resources.
%We therefore repeated all the runs in the table for the larger value
%$l=0.0025$, maintaining {\em fixed values} of the physical ratios
%$\ldv/l, \lvi/l$. (This necessitated dividing the densities by the
%factor $(0.00250/0.00156)^2$.) Accordingly, each corresponding run
%should have the same basic physics, but now with greater resolution at
%the grid scale, and at the expense of finite size effects setting in
%earlier.  Performing the same comparison ($l=0.00156$ versus
%$l=0.0025$) in the case of an applied shear will allow us to comment
%on finite size scaling in our inertia-free runs, for which we find no
%steady state free of finite size effects.

To study the effect of an applied shear flow, we consider the same
sets of parameter values as in zero shear but now with non-zero values
of $\gdot$. In anticipation of the anisotropy that shear will induce,
we also double the length of cell in the flow direction, along with
$N_x$ to maintain a constant density of mesh points.  See
tables~\ref{tab:thirdtable} and~\ref{tab:fourthtable}: as just
discussed, the parameters of any sheared run (R032s, for example) are
the same as those of the corresponding unsheared run (R032u), apart
from the values of $\gdot, \Lambda_x$ and $N_x$. The parameters of
runs R028s, R022s, R029s, R020s, R030s, R019s and R032s are those used
in the lattice Boltzmann study of Ref.~\cite{stansell-prl-96--2006},
converted into our units. The appearance of the new sets R028b, R022b
and R029b will be explained in Sec.~\ref{sec:inertia} below.

Assuming for the moment -- in some cases incorrectly, as we shall show
below -- that any sheared system will attain a steady state with a
typical domain size set by the reciprocal shear rate, our aim would
then be to construct scaling plots $L(1/\gdot)$ analogous to those of
$L(t)$ for the zero-shear coarsening regime in
Fig.~\ref{fig:domain_u}.  To obtain as much information as in
Fig.~\ref{fig:domain_u}, however, we would need to run for very many
values of the shear rate $\gdot$ to produce corresponding near
continuous segments of $L$ versus $1/\gdot$.  This is prohibitive
within available computing resources.  Accordingly, we ran for just
one shear rate for each set of (other) parameter values, marking the
scaled reciprocal shear rates recorded in tables~\ref{tab:thirdtable}
and~\ref{tab:fourthtable} with vertical arrows in
Fig.~\ref{fig:domain_u}.

\section{Numerical results}
\label{sec:results}

We now present the results of our numerical simulations. We start in
Sec.~\ref{sec:zeroShear} with the case of coarsening in zero shear,
before considering an applied shear flow with and without inertia in
Secs.~\ref{sec:inertia} and~\ref{sec:noInertia} respectively.
Sec.~\ref{sec:link} contains a linking discussion. The results in
Sec.~\ref{sec:noInertia} are (we believe) novel. In contrast, our aim
in Secs.~\ref{sec:zeroShear} and~\ref{sec:inertia} is simply to
reproduce behaviour seen earlier in the literature by lattice
Boltzmann
techniques~\cite{wagner-prl-80-1429-1998,kendon-jfm-440-147-2001,stansell-prl-96--2006}.
The purpose of this comparative part of our study is threefold: First,
and foremost, to develop confidence in our own algorithm and the code
via which it is implemented; second, to demonstrate the method used
here, which is potentially simpler and faster than lattice Boltzmann,
to be equally capable of capturing the physics of demixing, in 2D
at least; and third, to provide an independent check of some recent
results concerning nonequilibrium steady states under
shear~\cite{stansell-prl-96--2006}.

In each run, we take as an initial condition at each grid point a
value of $\phi$ chosen at random from a flat probability distribution
between $-0.01$ and $+0.01$. We checked for independence with respect
to this initial condition (not shown) the statistical properties of
the domain morphologies presented below. This independence holds as
long as many domains are present, making the system self-averaging.
Accordingly, for each set of parameters we give results below for a
single simulation run only.

\subsection{Zero shear}
\label{sec:zeroShear}

Following a temperature quench into the two phase regime, domains of
each phase form.  These progressively coarsen in size through the
action of surface tension, passing through growth regimes that are
successively dominated by diffusive, viscous and inertial dynamics.
Within these regimes, dimensional analysis predicts a functional
dependence of the characteristic domain size upon time of $L_{\rm
  D}\propto t^{1/3}$, $L_{\rm V} \propto t$ and $L_{\rm I}\propto
t^{2/3}$ respectively, as discussed in Sec.~\ref{sec:lengthscales}.
Our aim in this section is to investigate this behaviour numerically,
as done previously by lattice Boltzmann
methods~\cite{wagner-prl-80-1429-1998,kendon-jfm-440-147-2001}.

\begin{figure}[tb]
\subfigure{\includegraphics[scale=0.36]{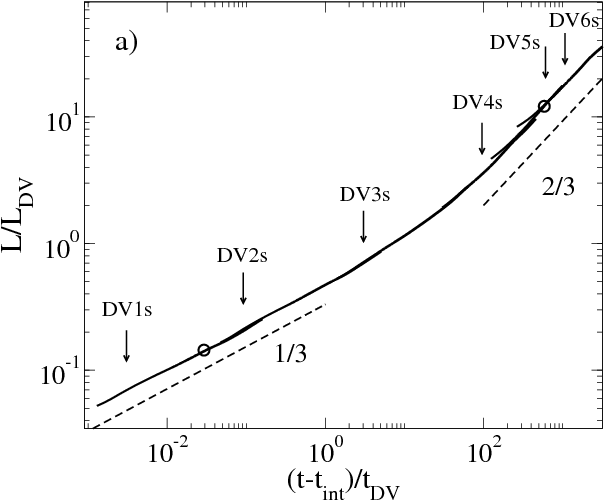}}
% Diffusive to viscous regime without shear.
% results/noShear/DVsets/DV{1,2,3,4,5,6}u_l0.00156_Dt0.0*_CG22
% Processed by scripts/domainSize with option 2, y, 2. Lower time
% cutoff 50.0. Upper domain cutoff (by this measure of structure
% factor) 0.015. Tdv for sets DV{1,2,3,4,5,6}u are respectively 33541,
% 1061, 33.54, 1.061, 0.174, 0.03354.  t_int/Tdv for sets
% DV{1,2,3,4,5,6}u are respectively 0.0003, 0.003, 0.17, 20.0, 180.0,
% ??  1/(gdot*Tdv) for sets DV{1,2,3,4,5,6}u are respectively 0.00298,
% 0.0943, 2.98, 94.3, 575, 994.
\hspace{0.4cm}
\subfigure{\includegraphics[scale=0.36]{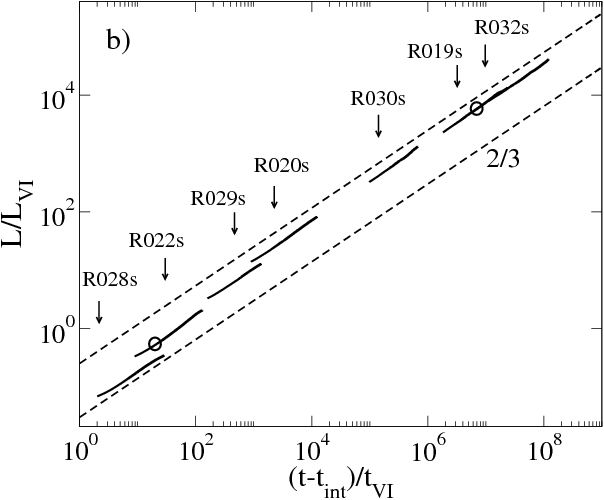}}
% Viscous to inertial regimes without shear. results files
% results/noShear/R0sets/R0*l0.00156*22. Processed by
% scripts/domainSize.pl with option 2, y, 2. Lower time cutoff of
% 40.0. Upper domain cutoff (by this measure of structure factor)
% 0.02. 
\caption{ a) Scaled domain size versus scaled time for coarsening
  in zero shear in the diffusive and viscous hydrodynamic regimes.
  Segments from left to right correspond to runs DV1u ... DV6u. The
  vertical arrows show the values of scaled reciprocal shear rate used
  in each corresponding run DV1s ... DV6s.  Circles show the times of
  the top two snapshots of Fig.~\ref{fig:state_u} below.
  \\\hspace{0.4cm} b) Scaled domain size versus scaled time for
  coarsening in zero shear in the viscous and inertial hydrodynamic
  regimes.  Segments from left to right correspond to runs R028u,
  R022u, R029u, R020u, R030u, R019u, R032u.  The vertical arrows show
  the values of scaled reciprocal shear rate used in each
  corresponding run R028s ...  R032s. Circles show the times of the
  bottom two snapshots of Fig.~\ref{fig:state_u} below.}
\label{fig:domain_u}
\end{figure}

\begin{figure}[tb]
\includegraphics[scale=0.6]{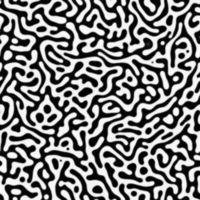}
\includegraphics[scale=0.6]{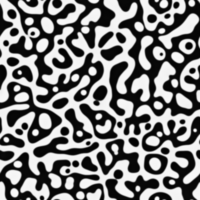}
\vspace{0.05cm}
\includegraphics[scale=0.6]{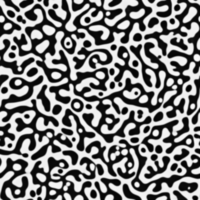}
\includegraphics[scale=0.6]{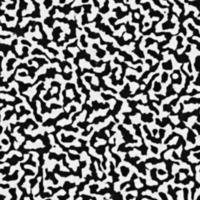}
\caption{Snapshots of domain morphology during coarsening in zero
  shear at the times shown by circles in Fig.~\ref{fig:domain_u}. Top
  left: diffusive regime run DV1u at $t=1052$. Top right: viscous
  hydrodynamic regime run DV5u at $t=131$. Bottom left: viscous
  hydrodynamic regime run R022 at $t=78.9$. Bottom right: inertial
  hydrodynamic regime run R019 at $t=78.9$.}
% Diffusive:
% results/noShear/DVsets/states/DV1u_l0.00156_Dt0.016_CG22_state.
% State at time 1052. (t-tint)/tdv is then 0.029 Viscous 1:
% results/noShear/DVsets/states/DV5u_l0.00156_Dt0.016_CG22_state State
% at time 131.6. (t-tint)/tdv is then 572.  Viscous 2:
% results/noShear/R0sets/states/R022_l0.00156_*22_state,
% state at time 78.96. At this time L=0.00775. For this set
% Lvi=0.0104. So L/Lvi=0.745.
% Inertial:
% results/noShear/R0sets/states/R019_l0.00156_*22_state,
% state at time 78.96. At this time L=0.0757. For this set
% Lvi=0.00000159. So L/Lvi=4761
\label{fig:state_u}
\end{figure}

As noted in Sec.~\ref{sec:parameters}, only a small segment of the
full coarsening curve $L(t)$ can be explored in any simulation run.
Accordingly, in different runs we vary the viscosity $\eta$ and
density $\rho$. In this way we correspondingly vary the crossover
length scales $\ldv$ and $\lvi$. We then combine the data sets from
the different runs into composite scaling curves of $L/\ldv$ and
$L/\lvi$, each spanning several decades in scaled length and time. We
choose as our measure of $L$ the one defined in
Eqn.~\ref{eqn:L_structure} via the structure factor.

To explore the diffusive and viscous hydrodynamic regimes, and the
crossover between them, we performed a single simulation run for each
set of parameter values of table~\ref{tab:secondtable}. The composite
curve of $L/\ldv$ is shown in Fig.~\ref{fig:domain_u}a).  For each
segment of the curve, three manipulations were performed. (i) The
first $t=0-40$ time units were discarded to allow for transient
dynamics during the initial process of domain formation. (ii) A
time-offset $t_{\rm int}$ was subtracted from the time $t$ to allow
for these non-universal dynamics during the initial transient. In each
case this subtraction was performed by eye.  The appropriate $t_{\rm
  int}$ was chosen as that which gives the most convincing straight
line on a log-log plot, without any known power present to bias the
eye during this process. (iii) The data set was cutoff at long times
once the typical domain size approaches the system size. For this
measure of the domain size, we used $L<0.02$ as a very conservative
criterion ensuring this. The same manipulations were performed for
runs in table~\ref{tab:firsttable}, to be described below.

For small values of $L/\ldv$, in the diffusive regime, we recover the
power $L_{\rm D} \propto t^{1/3}$ predicted by dimensional analysis.
This was confirmed previously by lattice Boltzmann
simulations~\cite{wagner-prl-80-1429-1998}. For large values of
$L/\ldv$, in the viscous hydrodynamic regime, dimensional analysis
predicts the growth law $L_{\rm V}\propto t$. We instead find $L
\propto t^{2/3}$ for each run.  This anomalous scaling been observed
previously, by lattice Boltzmann simulations. It is believed to stem
from a breakdown of scale invariance in this regime in 2D, due to the
formation of disconnected droplets~\cite{wagner-prl-80-1429-1998}.  As
a result of this breakdown, different measures of $L$ depend
differently on time (not shown here).  In 3D (not studied here) scale
invariance is recovered, along with the $2/3$ growth
exponent~\cite{kendon-jfm-440-147-2001}.

To study the inertial hydrodynamic regime and (again) the viscous
hydrodynamic regime, and the crossover between them, we use the
parameter values of table~\ref{tab:firsttable}. The composite curve of
$L/\lvi$ is shown in Fig.~\ref{fig:domain_u}b). For large values of
$L/\lvi$, in the inertial regime, we clearly recover the predicted
power $L_{\rm I} \propto t^{2/3}$. This has been seen before, in three
dimensions, by lattice Boltzmann
simulations~\cite{kendon-jfm-440-147-2001}.  For small values of
$L/\lvi$, in the viscous hydrodynamic regime, we again find the
anomalous power $L \propto t^{2/3}$ for each run. This is consistent
with the breakdown of scale invariance in 2D, and its attendant
departure from the predictions of dimensional
analysis~\cite{wagner-prl-80-1429-1998}.  Because of this breakdown,
segments R028, R022, R029 and R029 are not colinear: we do not have a
single composite scaling curve in this regime.  The fact that each
segment follows a $t^{2/3}$ power, apparently the same as in the
inertial regime, is a coincidence resulting from having chosen the
particular measure of domain size given by Eqn.~\ref{eqn:L_structure}.
As already noted, the lack of scaling in this regime results in
different power laws for different measures of the domain
size~\cite{wagner-prl-80-1429-1998}.

\subsection{Under shear, with inertia}
\label{sec:inertia}

We now turn to phase separation in the presence of an applied shear
flow. The main question of interest here is whether shear arrests
coarsening and restores a nonequilibrium steady state with a finite
domain size set by the reciprocal shear rate, independent of the
system size; or whether coarsening persists indefinitely, up to the
system size, even under shear.

Despite previous
experimental~\cite{hashimoto-jcp-88-5874-1988,takahashi-jr-38-699-1994,chan-pra-43-1826-1991,krall-prl-69-1963-1992,hashimoto-prl-74-126-1995,beysens-pra-28-2491-1983,chan-prl-61-412-1988,matsuzaka-prl-80-5441-1998,hobbie-pre-54-R5909-1996,qiu-pre-58-R1230-1998,lauger-prl-75-3576-1995},
numerical~\cite{shou-pre-61-R2200-2000,zhang-jcp-113-8348-2000,yamamoto-pre-59-3223-1999,corberi-prl-81-3852-1998,corberi-prl-83-4057-1999,corberi-pre-61-6621-2000,corberi-pre-62-8064-2000,rothman-prl-65-3305-1990,rothman-el-14-337-1991,chan-el-11-13-1990,qiu-jcp-108-9529-1998,padilla-jcp-106-2342-1997,ohta-jcp-93-2664-1990,wagner-pre-59-4366-1999,lamura-pamia-294-295-2001,lamura-epjb-24-251-2001,berthier-pre-6305--2001}
and
theoretical~\cite{onuki-jpm-9-6119-1997,lamura-pamia-294-295-2001,corberi-prl-81-3852-1998,corberi-prl-83-4057-1999,corberi-pre-61-6621-2000,corberi-pre-62-8064-2000,bray-ptrslsapes-361-781-2003,cavagna-pre-62-4702-2000,bray-jpag-33-L305-2000,Onuk97,doi-jcp-95-1242-1991}
work, this question remained open until the recent simulation study of
Stansell et al. in Ref.~\cite{stansell-prl-96--2006}.  Using lattice
Boltzmann techniques, they found nonequilibrium steady states of the
type reproduced by our own simulations in Fig.~\ref{fig:steadyStates}
(discussed below).  As expected under shear, the domain morphology is
anisotropic.  Accordingly, two length scales are needed to
characterise it.  The domain lengths $L_x$ and $L_y$ for the steady
states were extracted in Ref.~\cite{stansell-prl-96--2006} via the
curvature tensor of Eqn.~\ref{eqn:L_matrix}, and scaling plots of
$L_{\{x,y\}}/\lvi$ versus $1/\gdot \tvi$ were constructed. These
suggested apparent scaling exponents $L_x \sim \gdot^{-2/3}$ and
$L_y\sim \gdot^{-3/4}$, sustained over six decades. The same scalings
$L_{\|}\sim \gdot^{-2/3}$ and $L_{\bot}\sim \gdot^{-3/4}$ were
suggested for the major and minor principal lengths of the domains.
Slightly different scalings, discussed below, were obtained by the
same group in a more recent 3D study~\cite{cates-3d-2007}.

In this section we aim to show that our simulations, which are in 2D
throughout, reproduce these results to good approximation. We will
thereby gain confidence in our technique, before proceeding to the
novel contribution of this work in Sec.~\ref{sec:noInertia} below.
Accordingly, we perform a single simulation run for each set of
parameter values used in Ref.~\cite{stansell-prl-96--2006}, converted
into our units. See R028, R022, R029, R020, R030, R019 and R032 in
table~\ref{tab:thirdtable}.  The small discrepancy in the imposed
values of $1/\gdot \tvi$, evident in Fig.~\ref{fig:scalingPlot}, stems
from a slightly different definition adopted for the interfacial
width, realised by this author only at a late stage of the present
study.

\begin{figure}[tbp]
% results/shear/R0sets/R0{28b,22b,29b,30,20,19,32}*l0.00156*22 processed using scripts/domainSize.pl, with options 3 0 3 {l,s}
\subfigure{\includegraphics[scale=0.36]{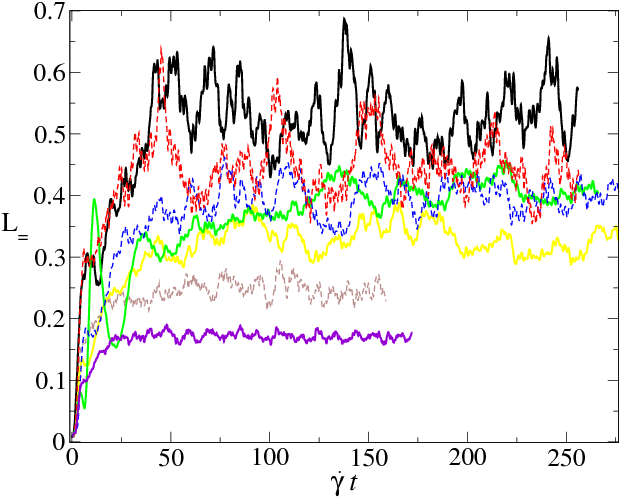}}
\subfigure{\includegraphics[scale=0.36]{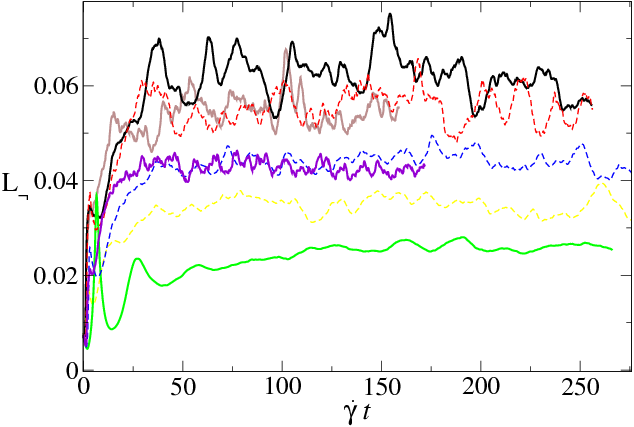}}
\caption{Top: larger domain length {\it vs.} strain $\gdot t$.  Data sets for
  decreasing values of long term temporal average correspond to R020s,
  R030s, R028b, R029bs, R022bs, R019s, R032s.  Bottom: smaller domain
  length {\it vs.} strain.  Data sets for decreasing values of long
  term temporal average correspond to R020s, R030s, R019s, R029bs,
  R032s, R022bs, R028b. Colour online.}
\label{fig:timeSeries}
\end{figure}

\begin{figure}[tbp]
% results/shear/R0sets/R0{28b,22b,29b,30,20,19,32}*l0.00156*22 processed using scripts/scalingDomainPlot.pl, with options 3 2 b and lower strain cutoff of 75
\subfigure{ \includegraphics[scale=0.38]{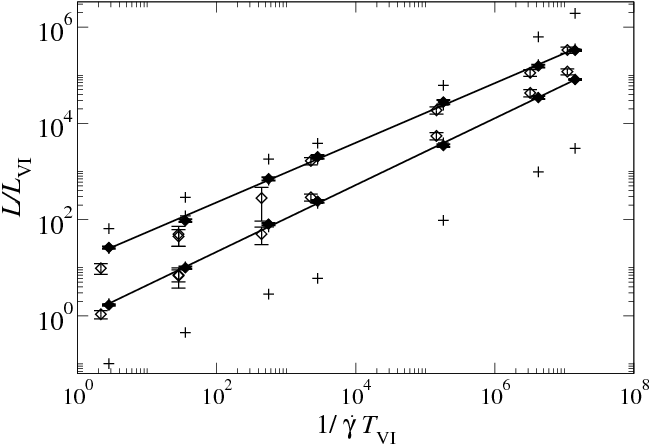}}
  \caption{Dimensionless scaling plot of lengths {\it vs.}  shear
    rate.  Solid symbols: average of time series of
    Fig.~\ref{fig:timeSeries} for $L_{\|}, L_{\bot}$ for strains
    $\gdot t>100$.  Symbols from left to right correspond to R028bs,
    R022bs, R029bs, R020s, R030s, R019s, R032s. Errors bars show the
    standard deviation.  Solid lines: power law fits to the solid
    symbols, suggesting $L_{\|} \sim \gdot^{-0.619}$ and $L_{\bot}\sim
    \gdot^{-0.693}$. Upper set of crosses shows the scaled system size
    $\Lambda_y/\lvi$; lower set shows the scaled interfacial width
    $l/\lvi$.  Open symbols: data from
    Ref.~\cite{stansell-prl-96--2006} for $L_{\|}, L_{\bot}$,
    reproduced here for comparison by kind permission of the authors
    of that study.}
\label{fig:scalingPlot}
\end{figure}

A typical run took one to two weeks of wall-clock time on a Linux box,
given exclusive use of a single 3.4GHz Intel Xeon CPU with 2Mb cache
and 400MHz DDR2 memory. Approximately $70\%$ of the runtime appears to
be used switching back and forth between real and reciprocal space at
each timestep.

In each run, we monitored as a function of time the characteristic
domain sizes $L_{\|}$ and $L_{\bot}$ extracted from the tensor of
Eqn.~\ref{eqn:L_matrix}. For each of R020, R030, R019 and R032 we
found $L_{\|}$ and $L_{\bot}$ to saturate at long times, showing
temporal fluctuations about constant mean values
(Fig.~\ref{fig:timeSeries}).  For the remainder of this section we
focus only on these statistically steady states, neglecting the first
$\gdot t=100$ strain units of each run.  (This cutoff was chosen by a
simple visual inspection of Fig.~\ref{fig:timeSeries}.) In each case,
finite size effects appear under control, as seen in the order
parameter snapshots of Fig.~\ref{fig:steadyStates}.  The mean values
of the time series are shown on the scaling plot of $L/\lvi$ versus
$1/\gdot\tvi$ in Fig.~\ref{fig:scalingPlot}, with error bars showing
the standard deviation. A snapshot of the order parameter for run R032
is shown in Fig.~\ref{fig:steadyStates} (bottom).

For runs R028, R022 and R029, in contrast, we found the domains to
wrap completely round the system in the flow direction, eventually
comprising trivial horizontal stripes connected at the edges of the
cell by the periodic boundary conditions. We believe this to be due to
the horizontal system size $\Lambda_x$ being dangerously close to the
expected values of $L_{\|}$ for these runs, leading eventually to
nucleation of these stripes.  

To eliminate this effect, we performed new runs R028b, R022b and R029b
at the same prescribed values of $1/\gdot \tvi$ as for R028, R022 and
R029, but now for larger values of the scaled system size
$\Lambda_x/\lvi$.  This was achieved by adjusting $\lvi$ at fixed
$\Lambda_x, \gdot\tvi$. Effectively, the left three sets of crosses in
Fig.~\ref{fig:scalingPlot} have been slightly shifted upward with
respect to those in Ref.~\cite{stansell-prl-96--2006} These new runs
produced nonequilibrium steady states, as seen in the time series of
Fig.~\ref{fig:timeSeries}. The long term time averages of these series
are shown in the scaling plot of Fig.~\ref{fig:scalingPlot}. A
snapshot of the order parameter for run R022b is shown in
Fig.~\ref{fig:steadyStates} (top).

Of course we cannot rule out eventual nucleation of system-wrapped
stripes at times exceeding those accessed numerically. The same
comment applies to runs R028, R029 and R022 of
Ref.~\cite{stansell-prl-96--2006}. Conversely, the system-wrapped
stripes seen in the corresponding runs in the recent 3D lattice
Boltzmann study~\cite{cates-3d-2007} might well be eliminated by
adjusting the scaled system size as done here.

% Reasonable quantitative agreement is found with the data of Ref. for
% $L_x$ and $L_y$. Complete agreement is not expected, since we
% instead give $L_{\|}$ and $L_{\bot}$. 

\begin{figure}[tbp]
\subfigure{ \includegraphics[scale=0.56]{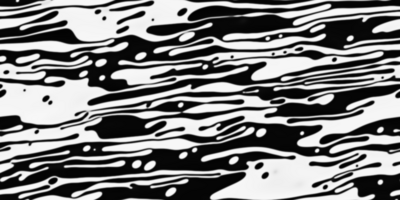}}
\subfigure{ \includegraphics[scale=0.56]{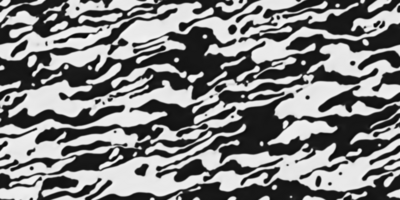}}
\caption{Snapshots of the steady state order parameter for R022b at strain $\gdot t=112$ (top) and R032 at strain $\gdot t=161$ (bottom).}
\label{fig:steadyStates}
\end{figure}

Power law fits to our data in the scaling plot of
Fig.~\ref{fig:scalingPlot} suggest exponents $L_{\|}\sim
\gdot^{-0.619\pm 0.01}$ and $L_{\bot}\sim ^{-0.693\pm 0.007}$. The
quoted uncertainties are the standard deviations given by the
automated regression package, and do not include systematic errors in
the simulations. In contrast Ref.~\cite{stansell-prl-96--2006} found
exponents for $L_{\|}, L_{\bot}$ of $-0.678\pm 0.039$ and $-0.756\pm
0.03$, by fitting to the open symbols that are reproduced by kind
permission in Fig.~\ref{fig:scalingPlot} for comparison with our data.
It also quoted exponents for $L_{x}, L_{y}$ of $-0.678\pm 0.042$ and
$-0.759\pm 0.029$.  The more recent 3D study~\cite{cates-3d-2007} by
the same group found exponents for $L_{\|}, L_{\bot}$ of $-0.64\pm
0.06$ and $-0.67\pm 0.03$, with $-0.53\pm0.04$ in the third dimension.
Within their $O(10\%)$ margins of error, these agree with the
exponents for $L_{\|}$ and $L_{\bot}$ found in the 2D study of the
present work.

\subsection{Discussion}
\label{sec:link}

In this section, we will discuss in more detail the shape of the
scaling plot of $L/\lvi$ versus $1/\gdot \tvi$ in
Fig.~\ref{fig:scalingPlot}.  In doing so, we will motivate the further
numerical study of Sec.~\ref{sec:noInertia} below.

Translated into equivalent scaled times $t/\tvi$, the range of scaled
shear rates $1/\gdot\tvi$ explored in Fig.~\ref{fig:scalingPlot} would
span the viscous and inertial hydrodynamic regimes in the coarsening
plot of Fig.~\ref{fig:domain_u}b).  One possibility for the shape of
the scaling curve $L/\lvi$ versus $1/\gdot\tvi$ under shear is that it
should be the same as that of the corresponding coarsening plot of
$L/\lvi$ versus $t/\tvi$ in zero shear, given the simple substitution
$t=\gdot^{-1}$. In view of this, it is instructive to compare
Fig.~\ref{fig:scalingPlot} with Fig.~\ref{fig:domain_u}b) in some
detail.

In the coarsening case (Fig.~\ref{fig:domain_u}b), two distinct
regimes are apparent. The inertial hydrodynamic regime, on the right
side, shows the $t^{2/3}$ exponent predicted by dimensional analysis.
Each segment neatly lines up with the next, consistent with dynamical
scaling.  In contrast, in the viscous hydrodynamic regime on the left
hand side the segments do not align.  Furthermore, each departs from
the predicted $t^1$ power.  As noted above, these anomalous features
stem from non-scaling effects that arise in 2D. In 3D these are
suppressed and the predicted $t^1$ power law is recovered.  The
counterpart of Fig.~\ref{fig:domain_u}b) for 3D systems thus comprises
a clean $t^1$ in the viscous regime on the left hand side, and
$t^{2/3}$ in the inertial regime on the right hand side, with a rather
slow crossover in between.  See Fig.  9 of
Ref.~\cite{kendon-jfm-440-147-2001}.  An applied shear flow is also
anticipated to suppress these non-scaling effects.  Accordingly, the
2D results of Fig.~\ref{fig:scalingPlot} are expected capture the
basic physics and, indeed, compare favourably with the recent 3D
sheared study of Ref.~\cite{kendon-jfm-440-147-2001}.

With these remarks in mind, it is now clear that we should in fact
compare Fig.~\ref{fig:scalingPlot} for sheared systems with Fig. 9 of
Ref.~\cite{kendon-jfm-440-147-2001} for coarsening systems.  As noted
above, one might naively expect the two plots to have the same shape,
up to the substitution $t=\gdot^{-1}$.

Instead, two differences are clearly apparent.  First,
Fig.~\ref{fig:scalingPlot} has two characteristic lengths, in contrast
to the single length that characterises unsheared systems in
Fig.~\ref{fig:domain_u}b).  With hindsight this is obvious: the domain
morphology is anisotropic in sheared systems, and so characterised by
two lengths. 

%The slightly different scaling with shear rate of $L_{\|}$ and
%$L_{\bot}$ suggests that these two lengthscales will eventually merge
%for large values of $1/\gdot\tvi$, and separate in the limit $1/\gdot
%\tvi \to 0$. In principle, this separation would lead to a crisis in
%which the domains assume an infinite aspect ratio.  This possibility
%is to be treated with some caution, however, because systematic
%uncertainties not accounted for in our data analysis do not rule out a
%common scaling for both lengths; and because a crisis-avoiding
%crossover might occur for values of $1/\gdot\tvi$ smaller than those
%accessed in Fig.~\ref{fig:scalingPlot}.

The second, and more subtle, difference is that sheared systems
apparently lack a distinction between viscous and inertial regimes,
with a single power law spanning all six decades in
Fig.~\ref{fig:scalingPlot}. A possible explanation for this is that
the range of scaled shear rates in Fig.~\ref{fig:scalingPlot} actually
occupies a window of extremely slow crossover between a true viscous
regime at smaller $1/\gdot\tvi$ and a true inertial regime at larger
$1/\gdot\tvi$, reminiscent of the slow crossover seen in the
coarsening plot in Ref.~\cite{kendon-jfm-440-147-2001}. Another is
that the value of $1/\gdot \tvi$ at which the crossover occurs in the
sheared case is much smaller than the corresponding crossover value of
$t/\tvi$ in the unsheared case, which is formally $O(1)$ but in
practice lies around $10^4$. The results in Fig.~\ref{fig:scalingPlot}
would then all lie in the inertial regime, to the right of this
crossover.

To test these ideas, in the next section we perform simulations at
strictly zero Reynolds number, $\rho=0$. We thereby take the limit
$\gdot \tvi \to \infty$ at the outset and perform simulations for
several finite values of $1/\gdot \tdv$. When converted into
equivalent scaled times $t/\tdv$ these span the diffusive and viscous
hydrodynamic regimes, as shown by the vertical arrows in
Fig.~\ref{fig:domain_u}a).  By analogy with the fact that the 3D
counterpart of the viscous regime of Fig.~\ref{fig:domain_u}a) would
be expected in the limit $t/\tdv \to\infty$ to match into that of the
viscous regime in Fig. 9 of Ref.~\cite{kendon-jfm-440-147-2001} in the
limit $t/\tvi\to 0$, we might then expect to access, for large
$1/\gdot \tdv \to\infty$, the scalings that $L_{\|}$ and $L_{\bot}$
would attain in the limit $1/\gdot\tvi \to 0$ at the extreme left of
an enlarged Fig.~\ref{fig:scalingPlot}.  

%Any discrepancy between these
%scalings and the scalings $L_{\|}\sim \gdot^{-0.619}, L_{\bot}\sim
%^{-0.693}$ already observed might indeed point to a slow crossover in
%Fig.~\ref{fig:scalingPlot}; or alternatively to a much smaller
%crossover value of $1/\gdot\tvi$ for sheared systems than of $t/\tvi$
%in the unsheared case.

Instead, however, we will find no evidence for non equilibrium steady
states in inertialess systems. In contrast, coarsening appears to
persist indefinitely, up to the system size, despite the applied shear
flow.  Given the existence of non equilibrium steady states for values
of $1/\gdot\tvi$ at the left hand edge of Fig.~\ref{fig:scalingPlot},
where the effects of inertia are non-zero but likely to be small, this
suggests that inertia plays the role of a singular perturbation in
this problem. Indeed, we will argue in Sec.~\ref{sec:argument} below
that the single scaling seen for each of $L_{\|}$ and $L_{\bot}$
across Fig.~\ref{fig:scalingPlot} results from a {\em mixed}
visco-inertial regime across the entire plot. The crisis of infinite
aspect ratio $L_{\|}/L_{\bot}\to \infty$ in the limit $1/\gdot/\tvi\to
0$, suggested by the data in Fig.~\ref{fig:scalingPlot}, is consistent
with our suggestion that inertia plays a singular role.

\begin{figure}[tbp]
\subfigure{\includegraphics[scale=0.36]{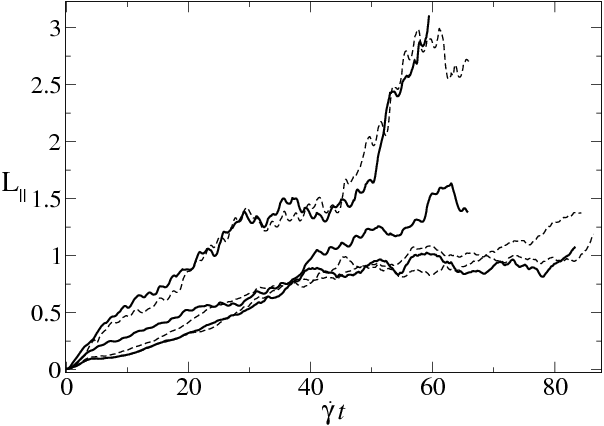}}
\subfigure{\includegraphics[scale=0.36]{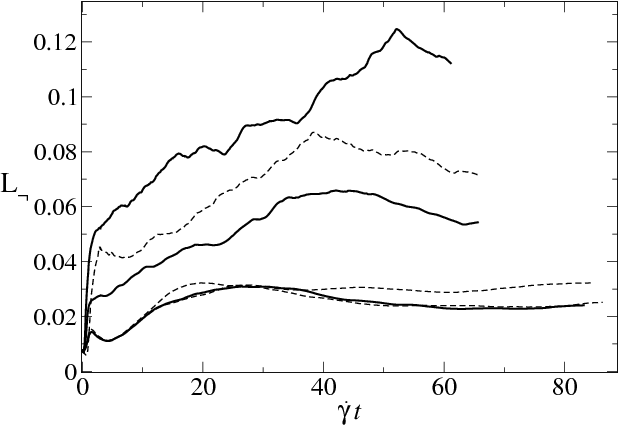}}
\caption{Top: larger domain length $L_{\|}$ {\it vs.} strain $\gdot
  t$.  Bottom: smaller domain length $L_{\bot}$ {\it vs.} strain
  $\gdot t$. In each case, decreasing values of $L$ at fixed $\gdot
  t=15$ correspond to runs DV5s, DV6s, DV4s, DV3s, DV2s, DV1s.}
\label{fig:timeSeries_zero_inertia}
\end{figure}

\begin{figure*}[tbp]
\subfigure[DV1s, $\gdot t=78.9$]{\includegraphics[scale=0.4]{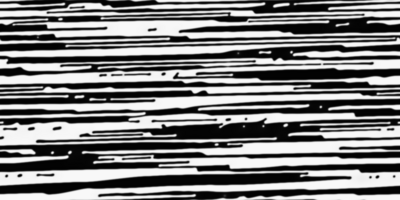}}
\subfigure[DV2s, $\gdot t=78.9$]{\includegraphics[scale=0.4]{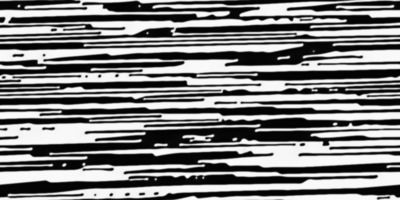}}
\subfigure[DV3s, $\gdot t=78.9$]{\includegraphics[scale=0.4]{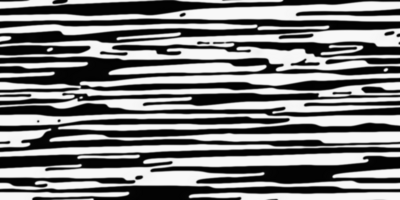}}
\subfigure[DV4s, $\gdot t=63.2$]{\includegraphics[scale=0.4]{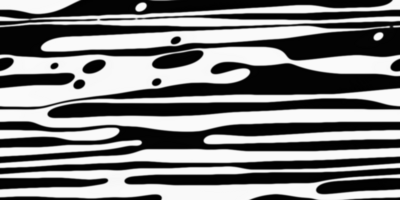}}
\subfigure[DV5s, $\gdot t=55.3$]{\includegraphics[scale=0.4]{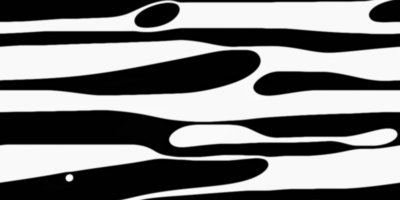}}
\subfigure[DV6s, $\gdot t=47.4$]{\includegraphics[scale=0.4]{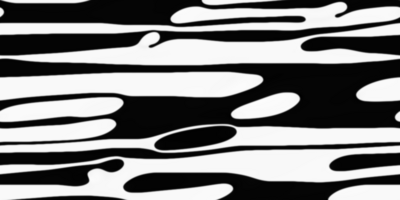}}
\caption{Snapshot domain morphologies after many strain units $S=\gdot
  t$ for reciprocal shear rates that, as equivalent times $\gdot^{-1}$,
  would be shown by the arrows in Fig.~\ref{fig:domain_u}a).
  These runs have zero inertia, $\rho=0$.}
\label{fig:spanning}
\end{figure*}

\subsection{Under shear, without inertia}
\label{sec:noInertia}

Motivated by the discussion of the previous section we now consider
phase separation under shear in inertialess systems, setting $\rho=0$.
As just discussed, by studying the limit $t/\tdv\to\infty$ we might
then, a priori, have expected to gain some insight into the far left
hand side of an enlarged Fig.~\ref{fig:scalingPlot}, $1/\gdot \tvi \to
0$.  Accordingly, we perform a single simulation run for each of the
parameter sets DV1s to DV6s in table~\ref{tab:fourthtable}. The values
of the scaled reciprocal shear rates $1/\gdot \tdv$ are marked as
corresponding scaled times $t/\tdv$ on the coarsening plot of
Fig.~\ref{fig:domain_u}a), where they are seen to span both the
diffusive and viscous hydrodynamic regimes.

In each run we monitored the characteristic domain sizes $L_{\|}$ and
$L_{\bot}$ as a function of time.  As can be seen in
Fig.~\ref{fig:timeSeries_zero_inertia}, in each case the larger length
$L_{\|}$ grows without bound until it attains the system size $O(1)$.
Correspondingly, a snapshot of the order parameter after many strain
units reveals system-wrapped domains, with pronounced finite-size
effects. See Fig.~\ref{fig:spanning}.  The snapshots for DV1s to DV3s
strongly resemble those reported in
Ref.~\cite{berthier-pre-6305--2001} for model B, which lacks
hydrodynamics. This is consistent with the fact that these runs occupy
the diffusive regime when marked as equivalent scaled times in
Fig.~\ref{fig:domain_u}b).

Beyond the parameter sets of table~\ref{tab:fourthtable}, we
furthermore performed runs for shear rates $\gdot=0.1, 0.03, 0.01,
0.003$ at each value of the viscosity $\eta=1000.0, 100.0, 10.0, 1.0,
0.3, 0.1$, thereby covering a complete rectangle in
$(\eta,\gdot)$-space, in contrast to the single slice taken by DV1s to
DV6s.  (For historical reasons these runs had a slightly larger value
$l=0.0025$ for the interfacial width, but we do not expect this to
make a qualitative difference.)  We found no evidence in any run for a
nonequilibrium steady state, unlimited by finite-size effects.

These results clearly suggest that the limit $1/\gdot\tdv\to\infty$,
which is approximated by runs DV5s and DV6s, does not match the limit
$1/\gdot\tvi\to 0$, which is approximated by runs R028b, R022b.  In
the next section, we propose that this is because the limit
$1/\gdot\tdv\to\infty$ corresponds to a pure viscous regime in which
no steady state exists; while the limit $1/\gdot\tvi\to 0$ corresponds to a
mixed visco-inertial regime in which a steady state does exist.

\section{Absence of nonequilibrium steady states in inertialess
  systems}
\label{sec:argument}

We now present an analytical argument in support of the above
numerical observation: that nonequilibrium steady states are not
attained in inertialess systems.  As we shall find, a closely related
argument supports the existence of a single power law for each of
$L_{\|}, L_{\bot}$ across the whole of Fig.~\ref{fig:scalingPlot}, as
seen first in Ref.~\cite{stansell-prl-96--2006}.  Recall that, a
priori, we might have expected this plot to comprise separate viscous
and inertial regimes.

As a preliminary step, we recall the case of domain coarsening in an
unsheared system. As discussed above, as time proceeds the system
passes through regimes that are successively dominated by diffusive,
viscous and inertial dynamics. In each regime, the system is aware of
the surface tension $\sigma$; the time $t$; and either the mobility
$M$ (diffusive regime), the viscosity $\eta$ (viscous regime) or the
density $\rho$ (inertial regime). In each case, there exists only one
possible combination of these relevant parameters that has the
dimensions of a length. Thus we have the following predicted growth
laws for the characteristic domain size:
\be
\label{eqn:coarsening}
L_{\rm D}=(M\sigma t)^{1/3},\;L_{\rm V}=\frac{\sigma t}{\eta},\;\;\textrm{and}\;\;L_{\rm I}=\left(\frac{\sigma t^2}{\rho}\right)^{1/3}.
\ee

Consider next a nonequilibrium steady state under shear. Neglecting
fluctuations, the time $t$ is now irrelevant. In its place, however,
we gain the reciprocal shear rate as a relevant parameter.  If
``pure'' diffusive, viscous and inertial regimes were still to exist,
then we could again only construct a {\em single} length scale to
characterise each:
\be
L_{\rm Ds}=(M\sigma \gdot^{-1})^{1/3},\;\;\;L_{\rm Vs}=\frac{\sigma \gdot^{-1}}{\eta},\;\;\textrm{and}\;\;L_{\rm Is}=\left(\frac{\sigma \gdot^{-2}}{\rho}\right)^{1/3},
\ee
by direct comparison with (\ref{eqn:coarsening}).

In contrast, however, our numerical results clearly show the domain
morphology to be anisotropic, as expected in a sheared system.  It is
therefore characterised by {\em two} lengths, which (according to
Fig.~\ref{fig:scalingPlot}) scale differently with the applied shear
rate. A steady state under shear therefore cannot exist in a ``pure''
diffusive, viscous or inertial regime, because in each there are
insufficient parameters out of which to construct the two lengthscales
needed to characterise it.

How can we retain enough parameters, out of the candidates $\sigma,
\gdot, M, \rho, \eta$ and $t$, to construct the two lengthscales
needed to characterise an anisotropic domain morphology under shear?
Assuming that $\sigma$ and $\gdot$ are always relevant, two options
are as follows.

\begin{enumerate}

\item 

  If the system is to exist in a {\em pure diffusive, viscous or
    inertial regime}, with knowledge of just one of $M, \eta$ or
  $\rho$ (as well as $\sigma$ and $\gdot$), it must retain dependence
  on time $t$.  Therefore, it must {\em fail to attain a steady
    state}.  

Assuming power laws,  a purely diffusive regime would then have 
\be
L_{\rm Ds}=(M \sigma t)^{1/3}(\gdot t)^a,
\ee
in which $a$ assumes two different values, which would be prescribed
by more detailed physics than is contained in the present argument.
Likewise, a purely viscous regime would have
\be
L_{\rm Vs}=\frac{\sigma t}{\eta}(\gdot t)^b,
\ee
again with two different values for $b$.

\item 

  If the system is to attain a {\em steady state}, thereby losing
  knowledge of the time $t$, it must retain dependence upon at least
  two of $M,\eta$ and $\rho$.  For example, a shear-induced steady
  state that is free of diffusion on the lengthscale of domains (no
  $M$) must exist in a {\em mixed viscous-inertial regime} with
  knowledge of both viscosity $\eta$ and density $\rho$:
\be
L_{\rm VIs}=\left(\rho^{-c-1}\eta^{c}\sigma \gdot^{-2-c}\right)^{\frac{1}{2c+3}},
\ee
with two different values for $c$.  

\end{enumerate}

We propose that both of these cases are seen in our numerical
simulations.  When inertia is present, for finite values of $\rho$
(however small), the system exists in a mixed visco-inertial steady
state, as discussed in option 2. This is consistent with the
observation of a single power law scaling for each of $L_{\|}$ and
$L_{\bot}$ across all six decades in the plot of $L/\lvi$ versus
$1/\gdot\tvi$ in Fig.~\ref{fig:scalingPlot} and in
Ref.~\cite{stansell-prl-96--2006}.  In contrast, when inertia is
strictly absent ($\rho=0$, infinite $\gdot\tvi$), the system exists in
a pure diffusive or viscous regime and never attains a steady state,
as in option 1.

To summarise: we suggest that the limit $1/\gdot\tdv\to\infty$
corresponds to a pure viscous regime with no steady state, while the
limit $1/\gdot\tvi\to 0$ corresponds to a mixed visco-inertial steady
state.  Thus we propose finally that inertia provides the role of a
singular perturbation in this problem.

\section{Summary and outlook}
\label{sec:conclusions}

We have studied numerically phase separation in binary fluids with
full hydrodynamics in two dimensions, considering both (1) unsheared
and (2) sheared systems, both (a) with and (b) without inertia.  Of
these, cases (1a), (1b) and (2a) have been studied by previous authors
using Lattice Boltzmann (LB)
methods~\cite{wagner-prl-80-1429-1998,wagner-pre-59-4366-1999,stansell-prl-96--2006,kendon-jfm-440-147-2001}.
In this paper, we have introduced an alternative simulation technique
that uses finite-differencing and spectral methods. We have also used
a convenient transformation to render trivial the implementation of
Lees-Edwards sheared periodic boundary
conditions~\cite{onuki-jpsj-66-1836-1997}.

In unsheared systems, phase separation occurs via a process of domain
coarsening.  Our simulation method successfully recovers results
obtained previously by LB for this process, both with and without
inertia (1a, 1b above).  In particular, it finds the familiar power
law $L_{\rm D}\sim t^{1/3}$ characterising the growth of the typical
domain size in the diffusive regime~\cite{wagner-prl-80-1429-1998}. It
also recovers $L_{\rm I}\sim t^{2/3}$ in the inertial hydrodynamic
regime~\cite{kendon-jfm-440-147-2001}. In the viscous hydrodynamic
regime it finds the anomalous power $L \sim t^{2/3}$, compared to the
predicted one of $L_{\rm v}\sim t^1$. As noted by previous authors,
this is due to subtle non scaling effects that arise in 2D from the
formation of disconnected droplets, also seen in LB
studies~\cite{wagner-prl-80-1429-1998}.  Such effects are eliminated
in 3D~\cite{kendon-jfm-440-147-2001} and also seem suppressed under
shear\cite{wagner-pre-59-4366-1999,stansell-prl-96--2006}.

We have also successfully reproduced the observations of existing LB
studies for sheared systems that have non-zero inertia (case 2a
above)~\cite{stansell-prl-96--2006}.  Here, an applied shear flow
arrests domain coarsening and restores a nonequilibrium steady state
with domains of a finite size set by the inverse shear rate.  The
domain morphology is anisotropic, characterised by two lengthscales
$L_{\|}, L_{\bot}$. Scaling exponents $L_{\|}\sim \gdot^{-0.619}$ and
$L_{\bot}\sim \gdot^{-0.693}$ found here agree with those of LB, to
within margins of error.  An outstanding puzzle, however, is why these
2D exponents agree better with the exponents $L_{\|}\sim
\gdot^{-0.64}, L_\bot \sim \gdot^{-0.67}$ of the 3D LB
study~\cite{cates-3d-2007}, than those of the 2D LB
study~\cite{stansell-prl-96--2006}, $L_{\|}\sim \gdot^{-0.678}$ and
$L_\bot\sim\gdot^{-0.756}$.  To investigate this, it would be
interesting to study the role of systematic errors in our simulations;
to consider the possible eventual nucleation of system wrapping
stripes in the 2D LB study of Ref.~\cite{stansell-prl-96--2006} for the
smaller values of $1/\gdot\tvi$; and to extend the present
finite-differencing work to 3D.

Our successful recovery of these important existing results in regimes
(1a, 1b, 2a) provides some confidence in our simulation method.
Beyond thereby having demonstrated this method to be capable of
capturing the physics of demixing, the other main contribution of this
paper has been a novel study of phase separation in sheared systems
that are strictly inertialess (case 2b).  Here we found no evidence of
nonequilibrium steady states, free of pronounced finite size effects.
Instead coarsening appears to persist indefinitely until the typical
domain size attains the system size, as in zero shear.

To support this observation, we have suggested by means of a simple
analytical argument that sheared inertialess systems adopt either a
pure diffusive or pure viscous regime, in each of which there are
insufficient parameters out of which to construct the two lengthscales
needed to characterise an anisotropic domain morphology in steady
state. By extending this argument slightly, we have also suggested
that sheared systems with any amount of inertia, however small, exist
in a mixed visco-inertial steady state.  This provides a possible
explanation for the observation of a single scaling with shear rate
for each of $L_{\|}$ and $L_{\bot}$ across all six decades in
Fig.~\ref{fig:scalingPlot}, and in the corresponding plot of
Ref.~\cite{stansell-prl-96--2006}

If this suggestion is correct, it remains unclear why the viscous and
inertial regimes should mix to yield a steady state, while the
diffusive and viscous regimes apparently remain separate, precluding
nonequilibrium steady states in truly inertialess systems. A possible
explanation lies in the more severe nonlinearity (in $\vecv{v}$) of
the inertial terms in the equation of motion. 

In future work, we aim to investigate whether the absence of
nonequilibrium steady states in inertialess systems persists in 3D.
In extending our method to 3D, several challenges are to be faced.  Of
these, the main ones appear to be contained in the basic Navier-Stokes
equations of incompressible fluid flow, and are not complicated
significantly by any additional order parameters~\cite{mixedLB}:
$\phi$ in this model.  However, relatively standard methods do exist
for finite-differencing the incompressible Navier-Stokes equations in
3D, as discussed in Ref.~\cite{Pozrikidis}.  At each timestep these
involve updating the vorticity via a slightly modified equation of
motion $\partial_t \omega=\cdots$; then updating the velocity field
either in the velocity-vorticity formulation (as effectively done here
in 2D via the intermediate of the streamfunction) or in the vector
potential-vorticity formulation.  In neither case does the pressure
field need to be calculated directly.  An outstanding issue before any
such algorithm could be run efficiently concerns its scaling with
system size, and its corresponding ease of parallelisation.  The same
question is also relevant in 2D, and a direct comparison of our
algorithm with that of Ref.~\cite{stansell-prl-96--2006} would clearly
be interesting.  In the longer term, the outcome of such studies might
help determine whether finite differencing can emerge as a useful tool
in such problems, alongside the already tested and reliable LB
methods.

Other open questions concern the role of initial conditions in sheared
systems. All the simulations reported here consider a temperature
quench performed in the presence of a shear flow. Future work should
consider an already demixed system, with either a flat interface or a
minimal-surface droplet, subsequently subject to a sudden shear
startup. We also aim to address demixing in complex macromolecular
fluids, focusing on the role of viscoelasticity.

\section*{Acknowledgements} 

The author thanks Prof. Alan Bray, Prof. Mike Cates and Prof. Peter
Sollich for helpful discussions and feedback, and the UK's EPSRC
GR/S29560/02 for funding.

\appendix

\bw

\section{Transformation to the cosheared frame}
\label{sec:appendix1}

Here we give details of the transformation to the cosheared
frame~\cite{onuki-jpsj-66-1836-1997}. As discussed in
Sec.~\ref{sec:numerics} above, this is performed in order to render
trivial the numerical implementation of sheared periodic boundary
conditions. As a first step, we separate the velocity field into a
constant affine contribution $\gdot y\xhat$ and a fluctuating part
$\vt$
\be
\vecv{v}(\vecv{x},t)=\gdot y\xhat + \vt(\vecv{x},t).
\ee
Noting that $\gdot y \xhat$ automatically satisfies the
incompressibility condition, we define the fluctuating parts of the
streamfunction and vorticity via
\be
\label{eqn:streamT}
\vt=\nablu\wedge \st \zhat,
\ee
and
\be
\label{eqn:streamVortT}
\wt=-\nabla^2\st.
\ee
Our equation set then comprises (\ref{eqn:streamT}) and
(\ref{eqn:streamVortT}) together with
\be
\label{eqn:vorticityT}
\rho\left(\partial_t\,\wt+\vt.\nablu\,\wt\right)+\rho \gdot y\, \partial_x\wt=\eta\nabla^2\wt-\left[\nablu\wedge\phi\nablu\mu\right]\cdot\zhat,
\ee
from Eqn.~\ref{eqn:vorticity},
\be
\label{eqn:phiT}
\partial_t\,\phi + \vt.\nablu\,\phi +\gdot y\,\partial_x\phi=l^2\nabla^2\mu,
\ee
from Eqn.~\ref{eqn:phi1}, and
\be
\label{eqn:muAppendix}
\mu= \phi(\phi^2-1)- l^2 \nabla^2\phi,
\ee
from Eqn.~\ref{eqn:mu1}, unchanged.

We then make a transformation to the cosheared frame
\be
\label{eqn:trans1}
(x,y,t)\to(x'=x-\gdot t y, y'=y,t'=t).
\ee
The various partial derivatives then become
\be
(\partial_x,\partial_y,\partial_t) = (\partial_{x'},-\gdot t\partial_{x'}+\partial_{y'},-\gdot y\partial_{x'}+\partial_{t'}).
\ee
Accordingly, we define the cosheared 2D gradient operator
\be
\nablu_c=\xhat\, \partial_{x'}+\yhat\,(-\gdot t\partial_{x'}+\partial_{y'}).
\ee
Finally, for any function $a$ we write
\be
\label{eqn:trans4}
a(x,y,t)=A(x',y',t').
\ee
Throughout we continue to work with velocity components
$\tilde{v}_x,\tilde{v}_y$ and not $\tilde{v}_{x'},\tilde{v}_{y'}$.

Inserting Eqn.~\ref{eqn:streamT} into the $\vt\cdot\nablu$ terms on
the LHS of Eqns.~\ref{eqn:vorticityT} and~\ref{eqn:phiT}, and
performing the transformation (\ref{eqn:trans1}) to (\ref{eqn:trans4})
on Eqns.~\ref{eqn:streamVortT} to~\ref{eqn:muAppendix}, we get the
equation set
\be
\tilde{\Omega}=-\nabla_c^2\tilde{\Psi},
\ee
together with
\be
\label{eqn:T2}
\rho\,\partial_{t'}\tilde{\Omega}+\rho\left(\partial_{y'}\tilde{\Psi}\partial_{x'}\tilde{\Omega}-\partial_{x'}\tilde{\Psi}\partial_{y'}\tilde{\Omega}\right)=\eta\,\nabla_c^2\tilde{\Omega}-\left[\partial_{x'}\Phi\partial_{y'} M-\partial_{y'}\Phi \partial_{x'} M\right],
\ee
\be
\label{eqn:T3}
\partial_{t'}\Phi+\left(\partial_{y'}\tilde{\Psi}\partial_{x'}\Phi-\partial_{x'}\tilde{\Psi}\partial_{y'}\Phi\right)=l^2\nabla_c^2M,
\ee
and
\be
M=\Phi(\Phi^2-1)-l^2\nabla_c^2\Phi.
\ee
In these $M$ represents upper case $\mu$, not mobility. A priori, the
bracketed expressions in Eqns.~\ref{eqn:T2} and~\ref{eqn:T3} contain
terms in the applied shear rate $\gdot$. However in each bracket these
are equal and opposite, and so cancel. For clarity we finally drop the
tildes and dashes, and revert from upper to lower case. The final
governing equations are then as summarised in Sec.~\ref{sec:numerics}
above.

\section{Numerical method}
\label{sec:appendix2}

Here we discuss the numerical algorithm used to study the dynamical
evolution of $\phi(\vecv{x},t),\omega(\vecv{x},t)$ and
$\psi(\vecv{x},t)$, as specified by Eqns.~\ref{eqn:equation1}
to~\ref{eqn:equation4} and~\ref{eqn:coshearedNabluFinal}.  Our basic
strategy is to step along a grid of time values $t^n=n\Delta t$ for
$n=1,2,3\cdots$.  Discretization with respect to time of any quantity
$f$ is denoted $f(t^n)=f^n$, or sometimes by $f|^n$. At each timestep,
we update $\phi^n, \psi^n, \omega^n \to \phi^{n+1}, \psi^{n+1},
\omega^{n+1}$ in three separate stages.  First, we update the
compositional order parameter $\phi^n \to \phi^{n+1}$ according to
Eqns.~\ref{eqn:equation3} and~\ref{eqn:equation4} with fixed, old
values of the stream-function $\psi^n$. We then update $\omega^n \to
\omega^{n+1}$ using Eqn.~\ref{eqn:equation2} at fixed
$\phi^{n+1},\psi^n$.  Finally we update the streamfunction
$\psi^n\to\psi^{n+1}$ using Eqn.~\ref{eqn:equation1} at fixed
$\omega^{n+1}$.

\begin{enumerate}

\item The update $\phi^n\to\phi^{n+1}$ using Eqns.~\ref{eqn:equation3}
  and~\ref{eqn:equation4} is performed in two successive partial
  updates. In the first we implement the advective term in
  Eqn.~\ref{eqn:equation3} to give $\phi^n \to \phi^{n+1/2}$. In the
  second we implement the diffusive term to give $\phi^{n+1/2} \to
  \phi^{n+1}$. The advective term is handled using an explicit Euler
  algorithm~\cite{PreTeuVetFla92}. Temporarily setting aside the issue
  of spatial discretization, this can be written
\be
\label{eqn:advective}
\phi^{n+1/2}(x,y)=\phi^n - \Delta t \; \left(\partial_{y}{\psi}^n\partial_{x}\phi^n-\partial_{x}{\psi}^n\partial_{y}\phi^n\right).  
\ee
This is then spatially discretized on a rectangular grid of
$(\Lambda_x N/\Lambda_y) \times N$ mesh points in real space $(x-y)$,
with constant mesh intervals $\Dx=\Dy=\Lambda_y/N$. Using indices
$i=1\cdots \Lambda_x N/\Lambda_y$ and $j=1\cdots N$, we denote any
discretized function $f(x_i,y_j)=f_{ij}$, or sometimes $f|_{ij}$.
Periodic boundary conditions are imposed by setting
$f_{i(-1)}=f_{i({N}-1)}$, $f_{i0}=f_{iN}$, $f_{i(N+1)}=f_{i1}$,
$f_{i({N}+2})=f_{i2}$, and similarly in the $x$ direction. The
derivatives of $\psi$ in Eqn.~\ref{eqn:advective} are discretized as
follows:
\be
\label{eqn:p1y}
\partial_x\psi|_{ij}^n=\frac{1}{2\Dx}\left[\psi_{(i+1)j}-\psi_{(i-1)j}\right],
\ee
with
\be
\partial_y\psi|_{ij}^n=\frac{1}{2\Dy}\left\{\left(1-\frac{s}{2}\right)\left(\psi_{i(j+1)}-\psi_{i(j-1)}\right) + \frac{s}{2}\left(\psi_{(i-2)(j+1)}-\psi_{(i+2)(j-1)}\right)\right\}\;\;\textrm{if}\;\;\; s\ge 0, 
\ee
and
\be
\partial_y\psi|_{ij}^n=\frac{1}{2\Dy}\left\{\left(1+\frac{s}{2}\right)\left(\psi_{i(j+1)}-\psi_{i(j-1)}\right) - \frac{s}{2}\left(\psi_{(i+2)(j+1)}-\psi_{(i-2)(j-1)}\right)\right\}\;\;\textrm{if}\;\;\; s<0. 
\ee
The derivatives of $\phi$ in Eqn.~\ref{eqn:advective} are discretized
in the same way.

%We verified
%that this gives the same results as using more sophisticated
%third-order upwinding~\cite{Pozrikidis}:
%%
%\be
%\partial_y\phi_{ij}^n=\frac{1}{6\Dy}\left[\phi^n_{i(j-2)} -6 \phi^n_{i(j-1)} +3
%  \phi^n_{ij} +2 \phi^n_{i(j+1)}
%\right]\;\;\;\textrm{if}\;\;\; v_y|^n_{ij}>0,
%\ee
%
%while
%
%\be
%\partial_y\phi_{ij}^n=\frac{1}{6\Dy}\left[-\phi^n_{i(j+2)} +6 \phi^n_{i(j+1)} -3
%  \phi^n_{ij} -2 \phi^n_{i(j-1)}
%\right]\;\;\;\textrm{if}\;\;\; v_y|^n_{ij}<0,
%\ee
%
%with analogous expressions for the derivative of $\phi$ with
%respect to $x$.

It then remains to implement the diffusive part of Eqn.~\ref{eqn:equation3}:
\be
\partial_t\phi=-l^2\,\nabla_c^2\phi-l^4\,\nabla_c^4\phi+l^2\,\nabla_c^2 f
\ee
with $f=\phi^3$. After calculating $f$ on our rectangular grid in real
space, this equation is handled in reciprocal space by taking fast
Fourier transforms $x\to q_x$ and $y\to q_y$ using a standard NAG
routine~\cite{NAG}. The transformation in each dimension generates a
real and an imaginary part, so for each mode $\vecv{q}=(q_x,q_y)$ we
need to consider a vector of the (transposed) form
$\vecv{\phi}^T=(\phi_{\rm rr}, \phi_{\rm ir}, \phi_{\rm ri}, \phi_{\rm
  ii})$ where subscript ``r'' denotes real part, and ``i'' imaginary.
The respective transforms $\tens{D}_2$ and $\tens{D}_4$ of the
operators $l^2\,\nabla_c^2$ and $l^4\,\nabla_c^4$ can easily be found
analytically:
\be
\tens{D}_2=l^2\left(
\begin{array}{cccc}
D & 0 & 0 & -\Delta \\
0 & D & \Delta & 0 \\
0 & \Delta & D & 0 \\
-\Delta & 0 & 0 & D 
\end{array}
\right)
\;\;\;\;\textrm{and}\;\;\;\;
\tens{D}_4=l^4\left(
\begin{array}{cccc}
D^2+\Delta^2 & 0 & 0 & -2D\Delta \\
0 & D^2+\Delta^2  & 2D\Delta & 0 \\
0 & 2D\Delta & D^2+\Delta^2 & 0 \\
-2D\Delta & 0 & 0 &  D^2+\Delta^2
\end{array}
\right),
\ee
in which 
\vspace{-0.4cm}
\be
D=-(aq_x^2+q_y^2),\;\;\;\Delta=bq_xq_y\;\;\;\textrm{with}\;\;\;a=1+[s(t)]^2,\;\;\;b=2s(t).
\ee
For each $\vecv{q}$-mode, we then have
\be
\partial_t\vecv{\phi}=-\tens{D}_2\cdot\vecv{\phi}-\tens{D}_4\cdot\vecv{\phi}+\tens{D}_2\cdot\vecv{f}.
\ee
To evolve this in time, we use an explicit Euler algorithm for the
first and third terms, and a semi-implicit Crank-Nicolson
algorithm for the second term. Thus we have
\be
\vecv{\phi}^{n+1}-\vecv{\phi}^{n+1/2}=-\tilde{\tens{D}}_2\cdot\vecv{\phi}^{n+1/2}-\tfrac{1}{2}\tilde{\tens{D}}_4\cdot(\vecv{\phi}^{n+1}+\vecv{\phi}^{n+1/2})+\tilde{\tens{D}}_2\cdot\vecv{f}^n,
\ee
in which $\tilde{\tens{D}}_m=\Dt\,\tens{D}_m$ for $m=2,4$. Rearranging gives finally
\be
\vecv{\phi}^{n+1}=\vecv{\phi}^{n+1/2} + (\tens{\delta}+\tfrac{1}{2}\tilde{\tens{D}}_4)^{-1}\cdot\left( - \tilde{\tens{D}}_2\cdot\vecv{\phi}^{n+1/2} - \tilde{\tens{D}}_4 \cdot\vecv{\phi}^{n+1/2} +\tilde{\tens{D}}_2\cdot\vecv{f}^n\right).
\ee

\item We now update $\omega^n \to \omega^{n+1}$ using
  Eqn.~\ref{eqn:equation2} at fixed $\phi^{n+1},\psi^n$.  The
  advective term on the LHS is updated in the same way as its
  counterpart in the $\phi$ equation above. To avoid inefficient
  multiple switching between real and Fourier space, this is in fact
  done at the same time as the corresponding update of $\phi$ in 1.
  above. (This reordering leaves the algorithm exactly unchanged.)
  We then update the RHS of Eqn.~\ref{eqn:equation2}.  Divided across
  by $\rho$, this reads:
\be
\label{eqn:vortNum}
\partial_t\,\omega=\nu\,\nabla_c^2 \omega+ g,
\ee
in which $\nu=\eta/\rho$ and
\be
\label{eqn:G}
g=-\frac{1}{\rho}\left[\partial_{x}\phi\,\partial_{y} \mu-\partial_{y}\phi \,\partial_{x} \mu\right].
\ee
As a first step, we calculate $g(x,y)$ using the newly updated
$\phi^{n+1}$ from 1. To do so, we first calculate
\be
\mu=\phi(\phi^2-1)-l^2(1+[s(t)]^2)\partial_x^2\phi - l^2\partial_y^2\phi + 2l^2s\partial_x\partial_y\phi.
\ee
At each grid point $i,j$, we spatially discretize the derivatives in
this expression according to
\be
\partial^2_x\phi|_{ij}=\frac{1}{(\Dx)^2}\left[\phi_{(i+1)j}-2\phi_{ij}+\phi_{(i-1)j}\right],
\ee
similarly for $\partial^2_y\phi$, and
\be
\partial_x\partial_y\phi|_{ij}=\frac{1}{4 \Dx\Dy}\left[\phi_{(i+1)(j+1)}-\phi_{(i+1)(j-1)}-\phi_{(i-1)(j+1)}+\phi_{(i-1)(j-1)}\right].
\ee
The first order derivatives of $\phi$ and $\mu$ with respect to $x$
and $y$ in~(\ref{eqn:G}) are then discretized as
in~(\ref{eqn:p1y}). 

Having calculated $g(x,y)$ in real space, we then Fourier transform
Eqn.~\ref{eqn:vortNum} to get
\be
\partial_t\,\vecv{\omega}=\tens{C}\cdot\vecv{\omega}+\vecv{g},
\ee
in which we have used the same vector/matrix notation as in 1 above,
with $\tens{C}=\nu \tens{D}_2/l^2$. To evolve this in time, we use a
semi-implicit Crank-Nicolson algorithm for the first term on the RHS
to get
\be
\label{eqn:forOmega}
\vecv{\omega}^{n+1}-\vecv{\omega}^n=\tfrac{1}{2}\tilde{\tens{C}}\cdot(\vecv{\omega}^n+\vecv{\omega}^{n+1})+\tilde{\vecv{g}}^{n+1}
\ee
with $\tilde{\tens{C}}=\Dt\, \tens{C}$ and $\tilde{\vecv{g}}=\Dt\,
\vecv{g}$. The superscript $n+1$ on the last term (\ref{eqn:forOmega})
serves to remind us that $g$ was calculated using the new $\phi^{n+1}$
from part 1. Rearranging, we get finally
\be
\vecv{\omega}^{n+1}=\vecv{\omega}^n+(\tens{\delta}-\tfrac{1}{2}\tilde{\tens{C}})^{-1}\cdot(\tilde{\tens{C}}\cdot\vecv{\omega}^n+\tilde{\vecv{g}}^{n+1}).
\ee

\item We finally update the streamfunction $\psi^n\to\psi^{n+1}$ using
  Eqn.~\ref{eqn:equation1} at fixed $\omega^{n+1}$. For each
  $\vecv{q}-$mode, we have
\be
\vecv{\psi}^{n+1}=-\tens{E}\cdot\vecv{\omega}^{n+1}\;\;\;\textrm{in which}\;\;\;\tens{E}=\frac{1}{D^2-\Delta^2}\left(
\begin{array}{cccc}
D & 0 & 0 & \Delta \\
0 & D & -\Delta & 0 \\
0 & -\Delta & D & 0 \\
\Delta & 0 & 0 & D 
\end{array}
\right).
\ee
\end{enumerate}
Finally, we transform all functions back to real space and return to
step 1 to start the next timestep.

\vspace{-0.4cm}

\subsection*{Algorithm at zero Reynolds number}

\vspace{-0.2cm}

The algorithm discussed so far is suited to non-zero values of the
fluid density $\rho$. At zero Reynolds number, with $\rho=0$,
Eqn.~\ref{eqn:equation1} and~\ref{eqn:equation2} of our basic equation
set combine to give the simpler equation
\be
\label{eqn:equationNew}
0=-\eta\,\nabla_c^4{\psi}-\left[\partial_{x}\phi\,\partial_{y} \mu-\partial_{y}\phi \,\partial_{x} \mu\right],
\ee
and we need no longer consider the vorticity field $\omega$.
Equations~\ref{eqn:equation3} and~\ref{eqn:equation4} remain
unchanged. Correspondingly, step 1 of our algorithm is also
unchanged. Steps 2 and 3 now combine to give
\be
\vecv{\psi}^{n+1}=\tens{E}.\vecv{g}^{n+1},
\ee
As above, we calculate $g$ in real space then take a Fourier transform. 
In Fourier space
\be
\tens{E}=\frac{1}{(D^2+\Delta^2)^2-(2D\Delta)^2}\left(
\begin{array}{cccc}
D^2+\Delta^2 & 0 & 0 & 2D\Delta \\
0 & D^2+\Delta^2  & -2D\Delta & 0 \\
0 & -2D\Delta & D^2+\Delta^2 & 0 \\
2D\Delta & 0 & 0 &  D^2+\Delta^2
\end{array}
\right).
\ee
After calculating $\psi$, we revert to real space to start the next
timestep.

\ew

%%%%%%%%%%%%%%%%%%%%%%%%%%%%%%%%%%%%%%%%%%%%%%%%%%%%%%%%%%%%%%%%%%%%%%%%%%%%%
%\bibliographystyle{prsty}
%\bibliography{ackerson,actin,ajdari,articles,banding,barham,Berret,berrportdecruppe,berthier,books,bray,callaghan,cates,chandcits,cook,crystal,crystal_theory,Decruppe,deGennes,demixing,dhont,dnatheory,elasticTurbulence,fielding,fischer,Fischer,flowcryst,fredrickson,furukawa,gelbart,gonnella,goveas,graham,Groisman,head,hebraud,helfrich,HinchRallison,hsiao,Kadoma,larson,LCtheory,leal,lerougeDecruppe,lifshitz,likhtman,line_tension,maffettone,malkus,master,mccoy,membs,Mexican,modelH,noirez,notes,olmsted,onions,onuki,otherRelated,pagonabarraga,phanThien,phd1,phd,pine,PineHu,pomeau,poon,pratt,psolutions,ramaswamy,recent,rheochaos,rheofolks,ryan,salmon,SalmonManneville,savedrecs.txt,schoot,semenov,sgrband,shaqfeh,sood,sriram,stein,sureshkumar,tanaka,vansaarloos,vorticity,Wang,warren,weitz,wilson,worms2,worms3,worms,yeomans,yuan,zubarev}

\begin{thebibliography}{10}

\bibitem{hohenberg77}
P.~C. Hohenberg and B.~I. Halperin, Rev. Mod. Phys. {\bf 49},  435  (1977).

\bibitem{bray-aip-43-357-1994}
A.~J. Bray, Adv. In Phys. {\bf 43},  357  (1994).

\bibitem{cates-fd--1-1999}
M.~E. Cates, V.~M. Kendon, P. Bladon, and J.~C. Desplat, Faraday Discussions
  {\bf 112},  1  (1999).

\bibitem{hashimoto-jcp-88-5874-1988}
T. Hashimoto, T. Takebe, and S. Suehiro, J. Chem. Phys. {\bf 88},  5874
  (1988).

\bibitem{takahashi-jr-38-699-1994}
Y. Takahashi, N. Kurashima, I. Noda, and M. Doi, J. Rheology {\bf 38},  699
  (1994).

\bibitem{chan-pra-43-1826-1991}
C.~K. Chan, F. Perrot, and D. Beysens, Phys. Rev. A {\bf 43},  1826  (1991).

\bibitem{krall-prl-69-1963-1992}
A.~H. Krall, J.~V. Sengers, and K. Hamano, Phys. Rev. Lett. {\bf 69},  1963
  (1992).

\bibitem{hashimoto-prl-74-126-1995}
T. Hashimoto, K. Matsuzaka, E. Moses, and A. Onuki, Phys. Rev. Lett. {\bf 74},
  126  (1995).

\bibitem{beysens-pra-28-2491-1983}
D. Beysens, M. Gbadamassi, and B. Moncefbouanz, Phys. Rev. A {\bf 28},  2491
  (1983).

\bibitem{chan-prl-61-412-1988}
C.~K. Chan, F. Perrot, and D. Beysens, Phys. Rev. Lett. {\bf 61},  412  (1988).

\bibitem{matsuzaka-prl-80-5441-1998}
K. Matsuzaka, T. Koga, and T. Hashimoto, Phys. Rev. Lett. {\bf 80},  5441
  (1998).

\bibitem{hobbie-pre-54-R5909-1996}
E.~K. Hobbie, S.~H. Kim, and C.~C. Han, Phys. Rev. E {\bf 54},  R5909  (1996).

\bibitem{qiu-pre-58-R1230-1998}
F. Qiu, J.~D. Ding, and Y.~L. Yang, Phys. Rev. E {\bf 58},  R1230  (1998).

\bibitem{lauger-prl-75-3576-1995}
J. Lauger, C. Laubner, and W. Gronski, Phys. Rev. Lett. {\bf 75},  3576
  (1995).

\bibitem{shou-pre-61-R2200-2000}
Z.~Y. Shou and A. Chakrabarti, Phys. Rev. E {\bf 61},  R2200  (2000).

\bibitem{zhang-jcp-113-8348-2000}
Z.~L. Zhang, H.~D. Zhang, and Y.~L. Yang, J. Chem. Phys. {\bf 113},  8348
  (2000).

\bibitem{yamamoto-pre-59-3223-1999}
R. Yamamoto and X.~C. Zeng, Phys. Rev. E {\bf 59},  3223  (1999).

\bibitem{corberi-prl-81-3852-1998}
F. Corberi, G. Gonnella, and A. Lamura, Phys. Rev. Lett. {\bf 81},  3852
  (1998).

\bibitem{corberi-prl-83-4057-1999}
F. Corberi, G. Gonnella, and A. Lamura, Phys. Rev. Lett. {\bf 83},  4057
  (1999).

\bibitem{corberi-pre-61-6621-2000}
F. Corberi, G. Gonnella, and A. Lamura, Phys. Rev. E {\bf 61},  6621  (2000).

\bibitem{corberi-pre-62-8064-2000}
F. Corberi, G. Gonnella, and A. Lamura, Phys. Rev. E {\bf 62},  8064  (2000).

\bibitem{rothman-prl-65-3305-1990}
D.~H. Rothman, Phys. Rev. Lett. {\bf 65},  3305  (1990).

\bibitem{rothman-el-14-337-1991}
D.~H. Rothman, Europhysics Lett. {\bf 14},  337  (1991).

\bibitem{chan-el-11-13-1990}
C.~K. Chan and L. Lin, Europhysics Lett. {\bf 11},  13  (1990).

\bibitem{qiu-jcp-108-9529-1998}
F. Qiu, H.~D. Zhang, and Y.~L. Yang, J. Chem. Phys. {\bf 108},  9529  (1998).

\bibitem{padilla-jcp-106-2342-1997}
P. Padilla and S. Toxvaerd, J. Chem. Phys. {\bf 106},  2342  (1997).

\bibitem{ohta-jcp-93-2664-1990}
T. Ohta, H. Nozaki, and M. Doi, J. Chem. Phys. {\bf 93},  2664  (1990).

\bibitem{wagner-pre-59-4366-1999}
A.~J. Wagner and J.~M. Yeomans, Phys. Rev. E {\bf 59},  4366  (1999).

\bibitem{lamura-pamia-294-295-2001}
A. Lamura and G. Gonnella, Physica A-statistical Mechanics Its Applications
  {\bf 294},  295  (2001).

\bibitem{lamura-epjb-24-251-2001}
A. Lamura, G. Gonnella, and F. Corberi, European Phys. J. B {\bf 24},  251
  (2001).

\bibitem{berthier-pre-6305--2001}
L. Berthier, L.~F. Cugliandolo, and J.~L. Iguain, Phys. Rev. E {\bf 6305},
  (2001).

\bibitem{onuki-jpm-9-6119-1997}
A. Onuki, J. Physics-condensed Matter {\bf 9},  6119  (1997).

\bibitem{bray-ptrslsapes-361-781-2003}
A.~J. Bray, Philosophical Transactions Royal Soc. London Series A-mathematical
  Phys. Engineering Sciences {\bf 361},  781  (2003).

\bibitem{cavagna-pre-62-4702-2000}
A. Cavagna, A.~J. Bray, and R.~D.~M. Travasso, Phys. Rev. E {\bf 62},  4702
  (2000).

\bibitem{bray-jpag-33-L305-2000}
A.~J. Bray and A. Cavagna, J. Phys. A-mathematical General {\bf 33},  L305
  (2000).

\bibitem{Onuk97}
A. Onuki, J. Phys. Cond. Matt. {\bf 9},  6119  (1997).

\bibitem{doi-jcp-95-1242-1991}
M. Doi and T. Ohta, J. Chem. Phys. {\bf 95},  1242  (1991).

\bibitem{stansell-prl-96--2006}
P. Stansell, K. Stratford, J.~C. Desplat, R. Adhikari, and M.~E. Cates, Phys.
  Rev. Lett. {\bf 96},    (2006).

\bibitem{cates-3d-2007}
K. Stratford, J.-C. Desplat, P. Stansell, and M.~E. Cates, Phys. Rev. E {\bf
  76},  030501  (2007).

\bibitem{onuki-jpsj-66-1836-1997}
A. Onuki, J. Phys. Soc. Japan {\bf 66},  1836  (1997).

\bibitem{wagner-prl-80-1429-1998}
A.~J. Wagner and J.~M. Yeomans, Phys. Rev. Lett. {\bf 80},  1429  (1998).

\bibitem{kendon-jfm-440-147-2001}
V.~M. Kendon, M.~E. Cates, I. Pagonabarraga, J.~C. Desplat, and P. Bladon, J.
  Fluid Mechanics {\bf 440},  147  (2001).

\bibitem{swift-pre-54-5041-1996}
M.~R. Swift, E. Orlandini, W.~R. Osborn, and J.~M. Yeomans, Phys. Rev. E {\bf
  54},  5041  (1996).

\bibitem{siggia}
E.~D. Siggia, Phys.~Rev. {\bf A20},  595  (1979).

\bibitem{furukawa-aip-34-703-1985}
H. Furukawa, Adv. In Phys. {\bf 34},  703  (1985).

\bibitem{kendon-prl-83-576-1999}
V.~M. Kendon, J.~C. Desplat, P. Bladon, and M.~E. Cates, Phys. Rev. Lett. {\bf
  83},  576  (1999).

\bibitem{pagonabarraga-jsp-107-39-2002}
I. Pagonabarraga, A.~J. Wagner, and M.~E. Cates, J. Statistical Phys. {\bf
  107},  39  (2002).

\bibitem{mixedLB} 
D. Marenduzzo, E. Orlandini, M.~E. Cates and J.~M.
  Yeomans, preprint arXiv:0708.2062 (2007).

\bibitem{Pozrikidis}
C. Pozrikidis, {\em Introduction to Theoretical and Computation Fluid Dynamics}
  (Oxford University Press, New York, 1997).

\bibitem{PreTeuVetFla92}
W.~H. Press, S.~A. Teukolsky, W.~T. Vetterling, and B.~P. Flannery, {\em
  Numerical Recipes in C (2nd ed.)} (Cambridge University Press, Cambridge,
  1992).

\bibitem{NAG}
Numerical Algorithms Group Ltd., Wilkinson House, Jordan Hill Road, Oxford, OX2
  8DR, UK.

\end{thebibliography}

\end{document}